\documentclass[12pt]{article}
\usepackage{booktabs}

\usepackage{epstopdf}% To incorporate .eps illustrations using PDFLaTeX, etc.
\usepackage[caption=false]{subfig}% Support for small, `sub' figures and tables
\usepackage{apacite}
\usepackage{color}
\usepackage[table]{xcolor}
\usepackage{mathrsfs}
\usepackage{amsfonts}
\usepackage{amsthm}
\usepackage{graphicx}
\usepackage{mathrsfs}
\usepackage{amsmath}
\usepackage{amssymb}
\usepackage{float}
\usepackage{subfig}
\usepackage{rotating,color}
\allowdisplaybreaks
\usepackage[authoryear]{natbib}
\usepackage{amsmath}
\usepackage{graphicx,psfrag,epsf}
\usepackage{enumerate,slashbox}
\usepackage{natbib}
\usepackage{url}
\usepackage{amssymb}
\usepackage{xr}
\usepackage{color}
\usepackage{setspace}
\usepackage{geometry}

\theoremstyle{plain}% Theorem-like structures provided by amsthm.sty

\newtheorem{Prop}{Proportion}[section]
\newtheorem{Lem}{Lemma}[section]
\newtheorem{Thm}{Theorem}[section]

\newenvironment{Prof}{\noindent\textup {\bf Proof\;}}{\hfill $\blacksquare$\par}

\def\bA{\boldsymbol{A}}\def\bB{\boldsymbol{B}}\def\bD{\boldsymbol{D}}

\def\bI{\boldsymbol{I}}

\def\bW{\boldsymbol{W}}\def\bX{\boldsymbol{X}}
\def\bY{\boldsymbol{Y}}\def\bZ{\boldsymbol{Z}}

\def\ba{\boldsymbol{a}}\def\bb{\boldsymbol{b}}
\def\bx{\boldsymbol{x}}
\def\by{\boldsymbol{y}}

\newcommand{\bmu}{\boldsymbol{\mu}}

\newcommand{\bepsilon}{\boldsymbol{\epsilon}}
\newcommand{\bgamma}{\boldsymbol{\gamma}}
\newcommand{\bdelta}{\boldsymbol{\delta}}
\newcommand{\bbeta}{\boldsymbol{\eta}}
\newcommand{\bxi}{\boldsymbol{\xi}}

\newcommand{\bGam}{\boldsymbol{\Gamma}}
\newcommand{\bSig}{\boldsymbol{\Sigma}}

\newcommand{\bOme}{\boldsymbol{\Omega}}
\newcommand{\bLam}{\boldsymbol{\Lambda}}

\def\b1{\boldsymbol{1}}
\def\b0{\boldsymbol{0}}

\newcommand{\R}{\mathbb{R}}
\newcommand{\N}{\mathbb{N}}

\newcommand{\E}{\mathbb{E}}

\newcommand{\Cov}{\operatorname{Cov}}

\newcommand{\Var}{\operatorname{Var}}
\newcommand{\var}{\operatorname{var}}

\newcommand{\Inf}{\mathrm{Inf}}

\newcommand{\diag}{\operatorname{diag}}

\newcommand{\tr}{\operatorname{tr}}

\newcommand*\Abs[1]{\left|#1\right|}    %   绝对值
\newcommand*\Braa[1]{\left(#1\right)}    %   大的括号
\newcommand*\Brab[1]{\left[#1\right]}    %   大的括号
\newcommand*\Brac[1]{\left\{#1\right\}}  %   大的大括号

\newcommand{\wideeq}[2][1.5]{
	\mathrel{\overset{#2}{\scalebox{#1}[1]{$=$}}}
}
\newcommand{\defeq}{\wideeq[1.5]{\triangle}}	%
\newcommand{\limP}{\xrightarrow{\ P\ }}	%

\newcommand{\limd}{\xrightarrow{\ \textup{d}\ }}

\begin{document}

%\articletype{ARTICLE TEMPLATE}% Specify the article type or omit as appropriate

\title{High dimensional Mean Test for Temporal Dependent Data}

\author{
Yuchen Hu and Xiaoyi Wang and Long Feng\\
Shanghai Jiaotong University; Beijing Normal University; Nankai University
}

\maketitle

\begin{abstract}
This paper proposes a novel test method for high-dimensional mean testing regard for the temporal dependent data. Comparison to existing methods, we establish the asymptotic normality of the test statistic without relying on restrictive assumptions, such as Gaussian distribution or M-dependence. Importantly, our theoretical framework holds potential for extension to other high-dimensional problems involving temporal dependent data. Additionally, our method offers significantly reduced computational complexity, making it more practical for large-scale applications. Simulation studies further demonstrate the computational advantages and performance improvements of our test.
\end{abstract}


\section{Introduction}

High-dimensional dependent data have gained increasing prominence in various disciplines, including finance, biomedical sciences, and geological studies. Such data typically exhibit both temporal dependence and correlations among different components, leading to intricate and multifaceted dependence structures. The complexity frequently renders traditional statistical approaches insufficient for effectively analyzing these data. This growing complexity has spurred the development of statistically valid inference procedures designed to address the challenges posed by high-dimensional dependent data. Among these challenges, the high-dimensional mean test is of particular importance.

In this paper, we focus on testing the mean vector of high-dimensional dependent data. Much of the existing literature has predominantly concentrated on methods developed for independent and identically distributed data, with the challenges posed by dependent data receiving comparatively less attention. Broadly speaking, these approaches can be categorized into five types, for a detailed review, see \citet{huang2022overview}. The first category includes modifications of the traditional Hotelling’s $T^2$  test, as explored by \citet{chen2011regularized, li2020adaptable} and others. Since the inversion of the covariance matrix becomes increasingly challenging in high-dimensional settings, alternative strategies have been developed. For example, \citet{bai1996effect} and \citet{chen2010two} examined the $L_2$ norm to measure the difference between mean vectors, leveraging largedimensional random matrix theory and U-statistic theory, respectively. Numerous other tests based on the $L_2$ norm have been proposed, including those by \citet{chen2019two, feng2016multivariate, gregory2015two, li2016simpler, srivastava2008test, wang2015high, zhong2013tests}. These tests are generally effective when signals are dense and relatively small but tend to lose power when dealing with sparse signals. To address the limitations of $L_2$ norm-based methods in detecting sparse and strong signals, $L_{\infty}$ norm-based methods have been proposed by \citet{tony2014two, chang2017simulation, xue2020distribution}. These methods provide higher power in scenarios where the alternative hypothesis involves sparse signals. Furthermore, adaptive methods that combine the strengths of both $L_2$ norm-based and $L_{\infty}$ norm-based approaches have been introduced by \citet{he2021asymptotically} and \citet{xu2016adaptive}. They proposed a relatively robust testing method by combining different $L_q$ norms, which is adaptable to alternative hypotheses with unknown levels of signal sparsity. Finally, the fifth category of methods involves random projections, as studied by \citet{srivastava2016raptt, thulin2014high} and others.

In contrast to the extensive research on independent data, relatively few studies have focused on testing high-dimensional means in temporally dependent data. Motivated by \citet{bai1996effect, ayyala2017mean} considers a M-dependent strictly stationary Gaussian process, and examined the $L_2$ norm of the sample average and investigated its asymptotic normality. In addition, \citet{zhang2023MeanTF} proposed a band-excluded U-statistic by removing the inner products between pairs of observations that are sufficiently distant from each other in the construction of the $L_2$ norm-based test statistic. They also established the asymptotic normality of this test statistic with $\beta$-mixing dependence. However, their method incurs a substantial computational cost, primarily due to the use of U-statistic-based estimators for estimating the asymptotic variance of the test statistic.

In this paper, we proposed a test statistic by removing all the inner products of the observations from the $L_2$ norm of the sample average. Compared to the method of \citet{ayyala2017mean}, our approach can enhance higher power, as previously demonstrated by \citet{chen2010two} under the independent setting. Furthermore, we demonstrate the asymptotic normality of the test statistic without relying on the stringent assumptions of Gaussian distribution or $M$-dependent structure among the observations. In our investigation of the theoretical properties, we assume that the data exhibit a linear dependence structure across observations. The primary contribution of this paper lies in the technical proof methodology. Importantly, our theoretical framework holds potential for extension to other high-dimensional problems involving temporal dependent data. Additionally, the computational burden of our method is significantly lower than that of \citet{zhang2023MeanTF}, as will be illustrated in the simulation studies in Section 3. Through this work, we aim to contribute to the understanding of mean testing in high-dimensional dependent data, offering a powerful and efficient approach that is better suited to handling the dependencies among components and observations.

The remainder of the paper is organized as follows. In section 2, we introduce the test statistic, and its asymptotic distribution under the null hypothesis. We conduct extensive simulation studies in Section 3. An empirical application is provided in Section 4. Section 5 contains some discussions. All technical details are included in the Appendix.

\section{Our Methods}\label{Sec 1}
\subsection{Test statistic}
%Let $\bZ_{t} = (Z_{t,1},\cdots, Z_{t,p})^\top$ be $p$-dimensional random vector, where $Z_{t,j}$ are independent and identically distributed random variables with $\E [Z_{t,j}] = 0$, $\E [Z_{t,j}^2] = 1$ and $\E [Z_{t,j}^4 ] = \mu_4 < \infty$. Define $Y_{t,j} = \sum_{k = 0}^{\infty} b_{k}Z_{t-k,j}$,
%\begin{align*}
%\bX_t = \bmu + \bSig^{1/2} \bY_{t}, \quad t =  1, \dots,  n,
%\end{align*}
%where $\bY_{t} = (Y_{t,1},\dots,Y_{t,p})^\top$.
We consider a $p$-dimensional stationary time series $\{\bX_t = (X_{t,1},\dots,X_{t,p})^\top\}_{t=1}^{n}$ with mean $\bmu$ and auto-covariance structure given by $\bGam_h = \Cov(\bX_t, \bX_{t+h})$ for $h=-(n-1),\dots,(n-1)$. We are interested in testing
\begin{align}\label{H0}
H_0: \bmu = \b0 \quad \text{versus} \quad H_1:  \bmu \neq \b0.
\end{align}
In this paper, we propose the following test statistic
\begin{align}
T_{n}= \frac12\bigg(n\bar{\bX}_n^\top \bar{\bX}_n - \frac1n\sum_{t = 1}^n \bX_t^\top \bX_t\bigg),
\end{align}
where $\bar{\bX}_n = n^{-1}\sum_{t = 1}^{n}\bX_t$. %It has been studied in many literatures about high dimensional location parameter testing problem designed for independent data, such as \cite{bai1996effect}, \cite{srivastava2009test}, \cite{chen2010two}, et al. When $\{\bX_t\}_{t=1}^{n}$  is  an $M$-dependent strictly stationary Gaussian process, the asymptotic normality of $n\bar{\bX}_n^\top \bar{\bX}_n$ has been established in \cite{ayyala2017mean}. Besides that, \cite{zhang2023MeanTF} introduced a U-statistic that effectively mitigates the influence of $\bX_t^\top\bX_s$ in scenarios where $|t-s|$ is large. They established the asymptotic normality of this test statistic under the assumption of $\beta$-mixing. However, their method entails a significant computational cost, primarily due to the employment of a U-statistic based estimator. Motivated by the important work of \cite{chen2010two}, we remove all diagonal terms $\bX_t^\top\bX_t$ from $n\bar{\bX}_n^\top \bar{\bX}_n$ for improving the convergence in high dimension framework. In this study, we demonstrate the asymptotic normality of $T_n$ by eliminating the stringent assumptions of Gaussian distribution and $M$-dependence. Our theoretical contribution has the potential to be extended to address the challenges associated with high-dimensional location testing. Furthermore, the computational burden of our method is significantly lower than that of \cite{zhang2023MeanTF}.
{
The problem of testing high-dimensional location parameters has been extensively studied for independent data, with notable contributions including \cite{bai1996effect}, \cite{srivastava2009test}, \cite{chen2010two}, et al. Recently, the literature has been extended to the high-dimensional settings, with works such as \cite{ayyala2017mean} and \cite{zhang2023MeanTF} proposing test statistics that estimate the $L_2$-type distance $\bmu^{\top}\bmu$ to quantify the discrepancy between the null and alternative hypotheses. Below, we provide a detailed mutual comparison of our proposed method with those of \cite{ayyala2017mean} and \cite{zhang2023MeanTF}.

\cite{ayyala2017mean} were among the first to address this problem in the context of high-dimensional dependent data. They assumed the data follows the high-dimensional $M$-dependent Gaussian stationary processes and proposed a test statistic based on the squared Euclidean norm of the sample mean vector, namely, $T_{\tiny{\mbox{APR}}} = \bar{\bX}_n^\top \bar{\bX}_n$. To address bias induced by the autocovariance structure, they constructed an unbiased estimator for $\E(\bar{\bX}_n^\top \bar{\bX}_n)$ and developed a ratio-consistent estimator for the asymptotic variance of the test statistic. Furthermore, they established the asymptotic normality under the null hypothesis as both sample size $n$ and the dimension $p$ tend to infinity, by assuming $p=O(n)$, $M=O(n^{1/8})$, and $\tr\{\bGam_a\bGam_b\bGam_c\bGam_d\} = o\{\tr^2(\bOme_n^2)\}$ for any $a,b,c,d=0,\dots,M$ , where $\bOme_n$ denotes the covariance matrix of $\bar{\bX}_n$ multiplied by $n$. A key limitation, however, is that their asymptotic theory relies heavily on the Gaussian process assumption. By relaxing the $M$-dependent Gaussian assumption, \cite{zhang2023MeanTF} introduced a band-excluded U-statistic (BEU) to estimate the $L_2$ norm of the population mean vector. Their test statistic is $T_{\tiny{\mbox{ZCQ}}} = {(n-b)(n-b+1)}^{-1}\sum_{|t_1-t_2|\geq b}\bX_{t_1}^\top\bX_{t_2}$, where $b$ is an exclusion bandwidth parameter. By systematically excluding cross-products of data vectors at temporally close points, their approach mitigates bias caused by long-range temporal dependencies. They studied the asymptotic theory of the BEU-statistic under a linear innovation model with $\beta$-mixing conditions. They considered two regimes of decay: exponential, which allows $p$ to grow exponentially with $n$, and polynomial, which requires $p = O(n^{\tau})$ for some $\tau > 0$. To estimate the asymptotic variance of the test statistic under the null hypothesis, they applied a similar band-exclusion technique and employed a kernel-smoothing technique with a quadratic spectral kernel, though this involves the substantial computational cost and nuisance bandwidth parameter.

%The method of \cite{ayyala2017mean} directly uses the squared Euclidean norm of the sample mean under the restrictive $M$-dependent Gaussian assumption.
Relative to \cite{ayyala2017mean}, we propose a test statistic by removing the terms of diagonal cross-product terms to enhance the power performance, which is inspired by \cite{chen2010two}. We adapt this insight to the dependent data framework. More fundamentally, our method diverges by relaxing the restrictive $M$-dependent Gaussian assumption to a more general linear process framework.
Compared to the method of \cite{zhang2023MeanTF}, the key difference lies in the estimation of the asymptotic variance. We  leverage a sample-splitting technique for the variance estimation of the asymptotic variance of the test statistic under a linear process framework. This fundamental difference circumvents the need for selecting the nuisance kernel smoothing parameter required in \cite{zhang2023MeanTF}. Furthermore, by departing from the U-statistic formulation, our method achieves a significant reduction in computational burden while maintaining theoretical consistency, rendering it more practical for large-scale applications.}
\subsection{Main results}
%Denote $\bGam_h \doteq \Cov(\bX_t, \bX_{t+h}) = a_{h}\bSig$ by the autocovariance matrix at lag $h$ of $\bX_t$ for $h=0,1,2,\dots$ and $\bGam_h=\bGam_{-h}^\top$, where $a_{h} \doteq \sum_{k = 0}^\infty b_kb_{k+h}$.
%and $\bOme = \bGam_0 + 2\sum_{h = 1}^\infty \bGam_h$.
Define $\bOme_{n}$, the covariance matrix of $\bar{\bX}_n$ multiplied by the sample size,
$$\bOme_{n} = \bGam_0 + 2\sum_{h = 1}^{n-1}\Big(1-\frac{h}n\Big)\bGam_h,$$
and $\bOme = \bGam_0 + 2\sum_{h = 1}^{\infty}\bGam_h$.
For investigating the asymptotic normality of $T_n$, we assume the following assumptions in the analysis.
\begin{description}
\item [Assumption 1.] For $t =  1, \dots, n$, $\bX_t = \bmu + \bSig^{1/2} \bY_{t}$, where $\bY_{t} = (Y_{t,1},\dots,Y_{t,p})^\top$ with $Y_{t,j} = \sum_{k = 0}^{\infty} b_{k}Z_{t-k,j}$, and $\{Z_{t,j}\}$ are independent and identically distributed (i.i.d.) random variables with $\E [Z_{t,j}] = 0$, $\E [Z_{t,j}^2] = 1$ and $\E [Z_{t,j}^4 ] = \mu_4 < \infty$.
\item [Assumption 2.]  There exist two positive constants $C_0$ and $C_1$ such that $\|\bSig\|_2 \leq C_0$ and $\tr(\bSig)/p \geq C_1$. Here $\|\cdot\|_2$ stands for the spectral norm of a matrix.
\item [Assumption 3.] (i) $b_n = o(n^{-5})$, i.e. $\lim\limits_{n\to \infty} n^{5}b_n = 0$; (ii) $\sum_{k = 0}^\infty b_k = s\neq 0$; (iii) $p = O(n)$ and $p \to \infty$ as $n\to \infty$.
%\item [Assumption 4.] Define $\bOme_{n}$, the covariance matrix of $\bar{\bX}_n$ multiplied by the sample size,
%$$\bOme_{n} = \bGam_0 + 2\sum_{h = 1}^{n-1}\Big(1-\frac{h}n\Big)\bGam_h\rightarrow \bF_0,$$
%where $\bF_0$ is the spectral matrix evaluated at the zero.
\end{description}
%\begin{itemize}
%\item [\textup{(A1)}] $p = O(n)$ and $p \to \infty$ as $n\to \infty$.
%\item[\textup{(A2)}] (i) $b_n = o(n^{-5})$, i.e. $\lim\limits_{n\to \infty} n^{5}b_n = 0$; (ii) $\sum_{i = 0}^\infty b_i = s\neq 0$.
%\item[\textup{(A3)}] $\sum_{i = 0}^\infty b_i = s\neq 0$.
%\item[\textup{(A3)}] There exist two positive constants $C_0$ and $C_1$ such that $\|\bSig\|_2 \leq C_0$ and $\tr(\bSig)/p \geq C_1$. Here $\|\cdot\|_2$ stands for the spectral norm of a matrix.
%\item [\textup{(A5)}] $x_0 = \sum_{k = 0}^\infty \tilde{b}_k\Sigma_1^{1/2}z_{-k} + \tilde{b}_{-1} \Sigma_{2}\tilde{z} + \tilde{b}_{-2}$. Where $\|\Sigma_1\|_2 \leq M_0$, $\|\Sigma_1\|_2 \leq M_0$. And $\tilde{z} = (\tilde{Z}_1,\cdots,\tilde{Z}_p)$ is independent of $z_t$ for any $t$, in which $\{\tilde{Z}_j\}$ are i.i.d. random variables with mean zero, variance one and finite forth moments. The coefficients satisfy $\sum_{k = -1}^\infty |\tilde{b}_k| = o(n^{1/2})$, and $\|\tilde{b}_{-2}\|^2 = O(p)$.
%\end{itemize}
Assumption 1 assume the linear process model of $\{\bX_t\}_{t=1}^n$, which is widely used in time series analysis (\citealt{Brockwell2013TimeST,zhang2018clt}). This linear process model cover the well known MA($q$) model and AR(1) model. Under the Assumption 1, we have $\bGam_h = a_{h}\bSig$, where $a_{h}=\sum_{k = 0}^\infty b_kb_{k+h}$. Assumption 2 implies that the correlation between the variables cannot be excessively large. Assumptions 3-(i) and 3-(ii) ensure that the temporal correlation becomes negligible as $n$ increases. Assumption 3-(iii) allows the dimension $p$ to increase to infinity as $n$ approaches infinity, which is a significant departure from the assumption that $n$ approaches infinity before $p$ does in \cite{breitung2005panel}.

Based on simple calculations, we have
$\E(T_{n}) = \sum_{h = 1}^{n-1}(1-\frac{h}{n}) \tr(\bGam_h)$ under $H_0$. Let $M = \lceil \min (n,p)^{1/8}\rceil$, we replace $\E(T_{n})$ by
\begin{align}
\label{Tn-mean}
\mu_n = \sum_{h = 1}^M \Big(1-\frac{h}{n}\Big)\tr(\bGam_h).
\end{align}
%The reasons for such substitution will be explained in Theorem \ref{Th 1}.
Due to the multiple occurrences of fourth moments, the variance of $T_n$ has a complicated structure. Define
%begin{align}\label{Tn-var}
%\sigma^2_n = \frac12\tr(\bOme^2) = \frac12\Big(\sum_{i = 0}^\infty b_i\Big)^4 \tr(\bSig^2).
%\end{align}
\begin{align}\label{Tn-var}
\sigma^2_n = \frac12\tr(\bOme^2).
\end{align}
To address this issue, we use $\sigma^2_n$ to approximate $\Var(T_n)$ under $H_0$.

\begin{Thm}\label{Th 1}
Under Assumptions 1--4 and $H_0$, if $M = \lceil \min (n,p)^{1/8}\rceil$, as $n \to \infty$,
\begin{align}
(T_n - \mu_n)/\sqrt{\sigma_n^2} \limd N(0,1),
\end{align}
where $ \mu_n$ and $\sigma_n$ is defined in \eqref{Tn-mean} and \eqref{Tn-var}, respectively.
\end{Thm}

%\subsection{Test procedure}
In order to construct the test statistic that can be used for the hypothesis test \eqref{H0},  we need to estimate $\mu_n$ and $\sigma_n^2$. We consider estimating $\tr(\bGam_h)$ with
\begin{eqnarray*}
\tr(\hat{\bGam}_h)=\frac{1}{n}\sum_{t=1}^{n-h} (\bX_t-\bar{\bX}_n)^\top (\bX_{t+h}-\bar{\bX}_n),
\end{eqnarray*}
and naturally propose the plug in estimator
\begin{eqnarray}
\hat{\mu}_n =  \frac{1}{n}\sum\limits_{h = 1}^{M}\sum\limits_{t = 1}^{n-h}\frac{n-h}{n}(\bX_t-\bar{\bX}_n)^\top (\bX_{t+h}-\bar{\bX}_n)
\end{eqnarray}
for $\mu_n$.
%\begin{Prop}\label{Prop for est of mu}
    %Take $M = \min\{n^{1/8},p^{1/8}\}$, then $\hat{\mu}_n$ is defined as follows,
    %\[\hat{\mu}_n = \sum_{h = 1}^M\frac{n-h}{n}\tr(\hat{\Gamma}_h) = \frac1n\sum_{h = 1}^{M}\sum_{t = 1}^{n-h}\frac{n-h}{n}(X_t-\bar{X}_n)^\top (X_{t+h}-\bar{X}_n)\]
    %Suppose that assumptions in (A1)-(A4) hold, and $H_0$ is true, $\hat{\mu}_n - \mu'_{n} = o_p(\tr^{1/2}(\Omega^2))$.
%\end{Prop}
We next find a ratio consistent estimator of $\sigma_n^2$. \cite{ayyala2017mean} introduced an unbiased estimator of $\sigma_n^2$. Their approach involves constructing a specific set of subscripts, ensuring that only pairwise correlations exist among the four samples in each component of the estimator. However, the consistency of their method is demonstrated under $M$-dependent Gaussian assumption. Building on this, \cite{zhang2018clt} propose a new estimator which is suitable for some special correlation structure. Motivated by \cite{zhang2018clt}, { we select the datapair $(\bX_t, \bX_s)$ that are sufficiently separated in time. Specifically, we require $|s-t|\geq \lfloor n/2\rfloor \gg M$, which make sure $\bX_t$ and $\bX_s$ are approximately independent under the assumed dependence structure. This allows us to use two non-overlapping segments of the series to estimate $\bGam_{h_1}$ and $\bGam_{h_2}$ independently. Thus, we propose}
\begin{eqnarray}
\label{est of sigma_T 1}
S_{h_1,h_2} = \frac{\displaystyle\sum_{t = 1}^{[n/2]-h_2}\displaystyle\sum_{s = t + [n/2]}^{n-h_2}
(\bX_{t}-\bar{\bX}_n)^\top(\bX_s-\bar{\bX}_n)(\bX_{t+h_1}-\bar{\bX}_n)^\top(\bX_{s+h_2}-\bar{\bX}_n)}{(n-h_2/2-\frac32 [n/2] +1/2)([n/2]-h_2)}.
\end{eqnarray}
to estimate $\tr(\bGam_{h_1}\bGam_{h_2})$.
It follows that
\begin{equation}\label{est of sigma_T 2}
\hat{\sigma}^2_n = \frac12\bigg(S_{0,0} + 2\sum_{h_1 = 1}^M S_{h_1,0} + 2\sum_{h_2 = 1}^M S_{0,h_2} + 4\sum_{h_1 = 1}^M\sum_{h_2 = 1}^M S_{h_1,h_2}\bigg).
\end{equation}
%\begin{Prop}\label{Prop for est of sigma}
    %Suppose that the conditions in Theorem \ref{Th 1} hold and that the estimator of $\sigma_n^2$ is as defined in \eqref{est of sigma_T 1} and \eqref{est of sigma_T 2}. Then as $\min\{n,p\}\to \infty $, $\hat{\sigma}^2_n/(\frac12\tr(\Omega^2))\limP 1$.
%\end{Prop}
\begin{Thm}\label{th22}
Under Assumptions 1--4 and $H_0$, if $M = \lceil \min (n,p)^{1/8}\rceil$, as $n \to \infty$,
\begin{align}
(T_n -  \hat{\mu}_n)/\sqrt{\hat{\sigma}^2_n} \limd N(0,1).
\end{align}
\end{Thm}

According to Theorem \ref{th22}, we reject the null hypothesis when  $(T_n - \hat{\mu}_n)/\sqrt{\hat{\sigma}^2_n}>z_{\alpha}$, where $z_\alpha$ is the upper $\alpha$-quantile of the standard normal distribution $N(0,1)$ for a level $\alpha$ test.

\section{Simulation Study}

In this section, we conduct a comparative analysis between our proposed method, referred to as $T_{n}$, and two established test approaches: the method developed by \cite{srivastava2009test} for independent data, denoted as $T_{\tiny\mbox{SRI}}$, and the approach introduced by \cite{zhang2023MeanTF} for $\beta$-mixing dependent data, denoted as $T_{\tiny\mbox{ZCQ}}$. We exclude the test proposed by \cite{ayyala2017mean} from this comparison due to the Gaussian $M$-dependent assumption, as well as its substantial computational demands.

We consider the following generating model
\begin{eqnarray}
\label{DGM}
\bX_{t}=\bmu+\bSig^{1/2}\sum\limits_{h=0}^q \bA_h\bZ_{t-h},
\end{eqnarray}
where $\bA_0,\dots,\bA_q$ are $p\times p$ symmetric matrices which determine the autocovariance structure of $\bX_{t}$, and
$Z_{t,j}$ are independently and identically distributed from $N(0,1)$ and $Gamma(4,2)-2$. %the uniform distribution over the interval $[-\sqrt{3},\sqrt{3}]$.
Besides that, we set $\bA_0=\bI_p$, and for  $h=1,2$,
\begin{equation*}
    \bA_{h,i,j} = \left\{\begin{array}{ll}
        \dfrac{\phi_1}{h}          & |i-j| = 0,\\
        \dfrac{\phi_1}{h|i-j|^2}  & 1\leq |i-j|\leq pw,\\
        0                                & |i-j|> pw,
    \end{array}\right.
         \quad\quad
    \bSig_{i,j} = \left\{\begin{array}{ll}
        1& |i-j| = 0,\\
        \dfrac{\phi_2}{|i-j|^2} & 1\leq |i-j|\leq pw,\\
        0& |i-j|> pw,
         \end{array}\right.
    \end{equation*}
where $(w,\phi_1,\phi_2)=(0.5,0.3,0.2)$. Here, $w$ represents the sparsity and $(\phi_1,\phi_2)$ control the strength of dependence between the variables. For $h>2$, we simply consider $\bA_{h,i,j}=e^{-2h}$.
In this paper, we choose $q = 0, 2, n-1$ in model (\ref{DGM}) to mimic the independent and dependent cases. To apply our proposed method effectively, we recommend taking $M=\lceil \min(n,p)^{1/8}\rceil$. For the comparison, we followed the recommendation outlined in \cite{zhang2023MeanTF} and take the lag window of their method as $\lfloor n/10 \rfloor$. We set the nominal significance level $\alpha = 0.05$, and all the simulation results are based on 1000 replications.

Table \ref{table1} summarizes the empirical sizes of the three tests under the null hypothesis. When $q=0$, indicating temporal independent, all testing procedures efficiently control the empirical type I error across a wide range of sample sizes and dimensions. However, under dependency setting, that is $q=2$ and $ q=n-1$, the $T_{\tiny\mbox{SRI}}$ test exhibits notable deviation from the nominal level, which is anticipated given their inability to account for the dependent structure. Compared to our method, the $T_{\tiny\mbox{ZCQ}}$ test shows a slight deviation from the nominal significance level in empirical type I error evaluation. Overall, our proposed test $T_n$ demonstrates superior robustness in maintaining the nominal size across different dependency scenarios.

\begin{table}[htbp]
\setlength{\abovecaptionskip}{0cm}
\setlength{\belowcaptionskip}{0.2cm}
\centering
    \caption{Empirical sizes (in \%) of different tests under different scenarios.}
\begin{tabular}{ccccccccccc}
   \toprule
   \multicolumn{2}{c}{}& \multicolumn{3}{c}{$q=0$} & \multicolumn{3}{c}{$q=2$} & \multicolumn{3}{c}{$q=n-1$} \\
   \cmidrule(r){3-5}\cmidrule(r){6-8}\cmidrule(r){9-11}
   $n$ & $p$ & $T_n$ & $T_{\mbox{\tiny{ZCQ}}}$ & $T_{\mbox{\tiny{SIR}}}$ & $T_n$ & $T_{\mbox{\tiny{ZCQ}}}$ & $T_{\mbox{\tiny{SIR}}}$ & $T_n$ & $T_{\mbox{\tiny{ZCQ}}}$ & $T_{\mbox{\tiny{SIR}}}$   \\
   \midrule
   \multicolumn{2}{c}{}& \multicolumn{9}{c}{$Z_{t,j}\sim N(0,1)$} \\
    \midrule
  250  & 100 & 7.5 & 5.8 & 3.8 & 6.7 & 6.5 & 98.4 & 5.8 & 6 & 98.3\\
          & 300 & 6.8 & 5.2 & 3.2 & 6.4 & 5.8 & 100 & 7.2 & 6.1 & 100\\
          & { 600} & { 6.7} & { 5.8} & { 1.4} & { 6.8} & { 4.3} & { 100} & { 7.3} & { 4.8} & { 100} \\
  500  & 100 & 6.5 & 5.6 & 4.8 & 7.2 & 7.4 & 98.8 & 7.2 & 7.3 & 98.7\\
          & 300 & 5.6 & 4.7 & 3.1 & 7.6 & 6.5 & 100 & 7.6 & 6.6 & 100\\
          &  { 600} &  { 6.4} &  { 4.7} &  { 2.3} &  { 7.6} &  { 6.8} &  { 100} &  { 7.2} &  { 6.4} &  { 100}\\
     \midrule
  \multicolumn{2}{c}{}& \multicolumn{9}{c}{$Z_{t,j}\sim Gamma(4,2)-2$} \\
  \midrule
  250  & 100 & 6.4 & 4.9 & 5 & 7 & 8 & 98.7 & 6.7 & 7 & 98.2\\
          & 300 & 5.6 & 4.5 & 2.8 & 5.6 & 5.1 & 100 & 8 & 6 & 100\\
          & { 600} & { 6.2} & { 4.9} & { 1.5} & { 6.9} & { 4.9} & { 100} & { 7.1} & { 4.6}  & { 100}\\
  500  & 100 & 5.6 & 4.8 & 4.6 & 6.2 & 6.4 & 98.6 & 7.6 & 7.7 & 98.9\\
          & 300 & 7 & 5.9 & 4.6 & 5.5 & 4.7 & 100 & 7.7 & 7.3 & 100\\
          &  { 600} &  { 7.1} &  { 5.9} &  { 3.6} &  { 6.7} &  { 6.7} &  { 100} &  { 7.3} &  { 5.8} &  { 100}\\
   \bottomrule
\end{tabular}
    \label{table1}
\end{table}

To compare statistical power, we examine six scenarios with different parameter configurations of $(n, p, q)$: { S1. $(n,p,q) = (250,300,0)$,  S2. $(n,p,q) = (250,300,2)$, S3. $(n,p,q) = (250,300,249)$, S4. $(n,p,q) = (250,100,0)$, S5. $(n,p,q) =(250,100,2)$ and S6. $(n,p,q) = (250,100,249)$. }
Set $\mu=(\omega_{Sk}\phi_3\mathbf{1}_{\lceil\nu p\rceil}^{\top},\mathbf{0}_{p-\lceil\nu p\rceil}^{\top})^{\top}$ with different scenarios of sparsity level $\nu \in \{0.2,0.4,0.8\}$ and signal strength $\phi_3 \in \{0.05,0.06,0.07,0.08, 0.09\}$, with $\omega_{S_1}=\omega_{S_4}=1$, $\omega_{S_2}=\omega_{S_5}=0.4$, $\omega_{S_3}=\omega_{S_6}=0.005$. %{ We consider six scenarios for different $(n, p, q)$, specifically: S1. $(n,p,q) = (250,100,0)$,  S2. $(n,p,q) = (250,100,2)$, S3. $(n,p,q) = (250,100,249)$, S4. $(n,p,q) = (250,300,0)$, S5. $(n,p,q) =(250,300,2)$ and S6. $(n,p,q) = (250,300,249)$. }
The outcomes are depicted in Tables~\ref{table3}--\ref{table6}. It’s evident that our proposed test, $T_{n}$, surpasses $T_{\mbox{\tiny{ZCQ}}}$ test. Furthermore, the computational cost of our proposed method is approximately one-tenth that of $T_{\mbox{\tiny{ZCQ}}}$ test, offering a significant advantage in terms of efficiency over the latter, which is presented in Table~\ref{table2}. This demonstrates that our method not only achieves better empirical power but also does so with a drastically reduced computational burden.

\begin{table}[htbp]
\setlength{\abovecaptionskip}{0cm}
\setlength{\belowcaptionskip}{0.2cm}
\setlength{\tabcolsep}{4pt}
\centering
    \caption{Timings (in seconds) of different tests under the null hypothesis with Gaussian distributed data, averaged over 1000 replications.}
\begin{tabular}{ccccccccccc}
   \toprule
   \multicolumn{2}{c}{}& \multicolumn{3}{c}{$q=0$} & \multicolumn{3}{c}{$q=2$} & \multicolumn{3}{c}{$q=n-1$} \\
   \cmidrule(r){3-5}\cmidrule(r){6-8}\cmidrule(r){9-11}
   $n$ & $p$ & $T_n$ & $T_{\mbox{\tiny{ZCQ}}}$ & $T_{\mbox{\tiny{SIR}}}$ & $T_n$ & $T_{\mbox{\tiny{ZCQ}}}$ & $T_{\mbox{\tiny{SIR}}}$ & $T_n$ & $T_{\mbox{\tiny{ZCQ}}}$ & $T_{\mbox{\tiny{SIR}}}$   \\
   \midrule
  250  & 100 & 0.780 & 27.865 & 0.004 & 0.753 & 26.220 & 0.004 & 0.716 & 26.342 & 0.004 \\
          & 300 & 1.293 & 72.004 & 0.047 & 1.265 & 73.016 & 0.048 & 1.275 & 71.536 & 0.046 \\
          & 600 & 8.985 & 357.492 & 1.311 & 9.859 & 358.353 & 1.571 & 8.079 & 352.216 & 1.174\\
  500  & 100 & 2.978 & 449.863 & 0.005 & 2.965 & 445.932 & 0.005 & 2.834 & 417.464 & 0.005 \\
          & 300 & 8.922 & 1986.630 & 0.061 & 8.522 & 2311.324 & 0.060 & 8.802 & 2283.405 & 0.059 \\
          & 600 & 26.122 & 2378.73 & 0.618 & 24.304 & 2515.59 & 0.597 & 26.083 & 2674.434 & 0.573\\
   \bottomrule
\end{tabular}
    \label{table2}
\end{table}

\begin{table}[htbp]
\setlength{\abovecaptionskip}{0cm}
\setlength{\belowcaptionskip}{0.2cm}
\centering
    \caption{Empirical powers (in \%) of different tests under scenarios S1--S3 with Gaussian distributed data.}
\begin{tabular}{ccccccccc}
   \toprule
   \multicolumn{2}{c}{}& \multicolumn{3}{c}{S1} & \multicolumn{2}{c}{S2} & \multicolumn{2}{c}{S3} \\
   \cmidrule(r){3-5}\cmidrule(r){6-7}\cmidrule(r){8-9}
   $\nu$ & $\phi_3$ & $T_n$ & $T_{\mbox{\tiny{ZCQ}}}$  &  $T_{\mbox{\tiny{SIR}}}$ & $T_n$ & $T_{\mbox{\tiny{ZCQ}}}$  & $T_n$ & $T_{\mbox{\tiny{ZCQ}}}$   \\
   \midrule
  0.2   & 0.05 & 48.6 & 43.9 & 29.6 & 15.9 & 14.3 & 18 & 15.7   \\
          & 0.06 & 69.4 & 66.1 & 52.3 & 21.2 & 18.9 & 24.1 &  20.8  \\
          & 0.07 & 85.5 & 83.1 & 74.5 & 28.6 & 25.5 & 32.2 & 28.3   \\
          & 0.08 & 95.3 & 94.3 & 89.6 & 38.3 & 34.1 & 41.7 & 37.8   \\
          & 0.09 & 98.8 & 98.7 & 97 & 48 & 44.8 & 52 & 48.1   \\
  0.4   & 0.05 & 46.6 & 41.6 & 29.9 & 28.1 & 24.8 & 32.8 & 29.1  \\
          & 0.06 & 69.2 & 65.2 & 51 & 41.3 & 38.3 & 45.5 &  41.6  \\
          & 0.07 & 85.2 & 82.6 & 73.9 & 57.7 & 54.9 & 60.4 & 57.1   \\
          & 0.08 & 94.9 & 94.1 & 89.4 & 71.4 & 67.7 & 73.9 & 70.3   \\
          & 0.09 & 98.7 & 98.3 & 96.4 & 83.3 & 81 & 85.4 & 84.1   \\
  0.8   & 0.05 & 48.9 & 44.3 & 30.9 & 58.3 & 53.5 & 59.2 & 55.9  \\
          & 0.06 & 68.8 & 65.4 & 53.1 & 79.6 & 76.8 & 78.8 & 75.3  \\
          & 0.07 & 85.6 & 83 & 74 & 91.4 & 89.5 & 90.6 & 89.6  \\
          & 0.08 & 95.5 & 94.2 & 90.2 & 97.2 & 96.9 & 97.6 & 96.6  \\
          & 0.09 & 99.2 & 99 & 97.1& 99.2 & 99 & 99.4 & 99.2  \\
  \bottomrule
\end{tabular}
    \label{table3}
\end{table}

\begin{table}[htbp]
\setlength{\abovecaptionskip}{0cm}
\setlength{\belowcaptionskip}{0.2cm}
\centering
    \caption{Empirical powers (in \%) of different tests under scenarios S4--S6 with Gaussian distributed data.}
{
\begin{tabular}{ccccccccc}
   \toprule
   \multicolumn{2}{c}{}& \multicolumn{3}{c}{S4} & \multicolumn{2}{c}{S5} & \multicolumn{2}{c}{S6} \\
   \cmidrule(r){3-5}\cmidrule(r){6-7}\cmidrule(r){8-9}
   $\nu$ & $\phi_3$ & $T_n$ & $T_{\mbox{\tiny{ZCQ}}}$  &  $T_{\mbox{\tiny{SIR}}}$ & $T_n$ & $T_{\mbox{\tiny{ZCQ}}}$  & $T_n$ & $T_{\mbox{\tiny{ZCQ}}}$   \\
   \midrule
  0.2   & 0.05 & 27.1 & 24.5 & 20.6 & 11.8 & 12 & 12.2 & 11.8   \\
          & 0.06 & 37.5 & 34.3 & 31 & 14.5 & 14.9 & 15.5 &  14.8  \\
          & 0.07 & 51.7 & 48.3 & 44.4 & 17.7 & 17.7 & 19.8 & 18.9   \\
          & 0.08 & 66.5 & 64.9 & 60.4 & 22.1 & 21.3 & 25.3 & 23.9   \\
          & 0.09 & 78.2 & 76.9 & 73.7 & 26 & 26 & 29.9 & 29.1   \\
  0.4   & 0.05 & 53.8 & 50.4 & 44.9 & 17.8 & 18 & 20 & 19.4   \\
          & 0.06 & 72.8 & 70 & 66.6 & 24 & 23.8 & 28 & 27.1   \\
          & 0.07 & 87.4 & 86.2 & 84.3 & 32.5 & 31.3 & 36.7 & 34.9   \\
          & 0.08 & 95.3 & 94.2 & 92.7 & 42.8 & 40.4 & 44.1 & 43.5   \\
          & 0.09 & 98.9 & 98.7 & 98.1 & 51.2 & 51.9 & 52.9 & 52.9   \\
  0.8   & 0.05 & 87.5 & 86.1 & 84.3 & 33 & 32.5 & 34.7 & 34.1   \\
          & 0.06 & 97.2 & 96.7 & 96.5 & 46.3 & 45.9 & 48.7 & 48.7   \\
          & 0.07 & 100 & 100 & 100 & 60.2 & 59.7 & 62.4 & 62.4   \\
          & 0.08 & 100 & 100 & 100 & 73.6 & 73.1 & 75.4 & 74.9   \\
          & 0.09 & 100 & 100 & 100 & 83.7 & 83.1 & 87.2 & 86.4   \\
  \bottomrule
\end{tabular}}
    \label{table4}
\end{table}

\begin{table}[htbp]
\setlength{\abovecaptionskip}{0cm}
\setlength{\belowcaptionskip}{0.2cm}
\centering
    \caption{Empirical powers (in \%) of different tests under scenarios S1--S3 with Gamma distributed data.}
\begin{tabular}{ccccccccc}
   \toprule
   \multicolumn{2}{c}{}& \multicolumn{3}{c}{S1} & \multicolumn{2}{c}{S2} & \multicolumn{2}{c}{S3} \\
   \cmidrule(r){3-5}\cmidrule(r){6-7}\cmidrule(r){8-9}
   $\nu$ & $\phi_3$ & $T_n$ & $T_{\mbox{\tiny{ZCQ}}}$  &  $T_{\mbox{\tiny{SIR}}}$ & $T_n$ & $T_{\mbox{\tiny{ZCQ}}}$  & $T_n$ & $T_{\mbox{\tiny{ZCQ}}}$   \\
   \midrule
  0.2   & 0.05 & 45.7 & 41.9 & 26.4 & 15.5 & 13 & 17.3 & 15.2  \\
          & 0.06 & 67 & 62.7 & 45.5 & 20.9 & 18.6 & 25.1 & 20.6  \\
          & 0.07 & 83.7 & 81.3 & 66.5 & 27.9 & 24.8 & 31.9 & 28.6  \\
          & 0.08 & 94.9 & 92.9 & 85.3 & 35.3 & 31.9 & 42.2 & 38.4  \\
          & 0.09 & 98.8 & 98.4 & 95.7 & 44.2 & 41.6 & 51.9 & 47.7  \\
  0.4   & 0.05 & 46.2 & 41.5 & 22.1 & 27.7 & 24.7 & 33 & 28.7  \\
          & 0.06 & 65.7 & 61.6 & 41.3 & 41.9 & 38.6 & 46.1 & 42  \\
          & 0.07 & 81.7 & 79.3 & 61.8 & 55.8 & 52.3 & 62 & 58.9  \\
          & 0.08 & 93.1 & 92.2 & 80.2 & 70.7 & 66.4 & 74.5 & 71.3  \\
          & 0.09 & 99 & 98.3 & 93.5 & 82.2 & 79.9 & 85 & 82.7  \\
  0.8   & 0.05 & 47.3 & 42.8 & 19.8 & 56.4 & 53 & 59 & 55.7  \\
          & 0.06 & 67.8 & 64.2 & 36.7 & 76.9 & 73.8 & 77.1 & 74.7  \\
          & 0.07 & 84.8 & 82.2 & 58 & 91.1 & 89.1 & 90 & 88.1  \\
          & 0.08 & 95.6 & 94.6 & 79.4 & 98 & 97.6 & 96.6 & 95.7  \\
          & 0.09 & 98.6 & 98.3 & 93.7 & 99.9 & 99.7 & 99 & 98.8  \\
   \bottomrule
\end{tabular}
    \label{table5}
\end{table}

\begin{table}[htbp]
\setlength{\abovecaptionskip}{0cm}
\setlength{\belowcaptionskip}{0.2cm}
\centering
    \caption{Empirical powers (in \%) of different tests under scenarios S4--S6 with Gamma distributed data.}
    {
\begin{tabular}{ccccccccc}
   \toprule
   \multicolumn{2}{c}{}& \multicolumn{3}{c}{S4} & \multicolumn{2}{c}{S5} & \multicolumn{2}{c}{S6} \\
   \cmidrule(r){3-5}\cmidrule(r){6-7}\cmidrule(r){8-9}
   $\nu$ & $\phi_3$ & $T_n$ & $T_{\mbox{\tiny{ZCQ}}}$  &  $T_{\mbox{\tiny{SIR}}}$ & $T_n$ & $T_{\mbox{\tiny{ZCQ}}}$  & $T_n$ & $T_{\mbox{\tiny{ZCQ}}}$   \\
   \midrule
  0.2   & 0.05 & 24.6 &  21.3 & 16.2 & 11.4 & 11.4 & 13.3 & 12.6   \\
          & 0.06 & 34.7 & 32.3 & 24.8 & 13.4 & 13.6 & 15.9 & 15.1   \\
          & 0.07 & 47 & 44.7 & 36.5 & 16.6 & 16.9 & 19.5 & 18.9   \\
          & 0.08 & 62.5 & 59.5 & 52.1 & 20.1 & 20.2 & 23.6 & 23.6   \\
          & 0.09 & 75.7 & 73.6 & 66.4 & 25 & 24 & 28.5 & 27.4   \\
  0.4   & 0.05 & 51.4 & 47.3 & 36.9 & 16.6 & 16.8 & 19.6 & 19.1   \\
          & 0.06 & 69.9 & 67.7 & 57.6 & 23 & 23.1 & 26.4 & 25.3   \\
          & 0.07 & 86.8 & 86.1 & 78.2 & 31.4 & 30.8 & 35.8 & 35.2   \\
          & 0.08 & 95.3 & 94.6 & 91.8 & 39.6 & 39.3 & 44.1 & 43   \\
          & 0.09 & 98.9 & 98.7 & 97.9 & 49.9 & 49.6 & 54.1 & 53.2   \\
  0.8   & 0.05 & 87.2 & 86.2 & 75.9 & 33.1 & 32.5 & 37.3 & 36.3   \\
          & 0.06 & 97.9 & 97.3 & 95.2 & 46.6 & 45.8 & 50.5 &  49.6  \\
          & 0.07 & 100 & 99.9 & 99.4 & 59.8 & 58.4 & 62.8 & 62.6   \\
          & 0.08 & 100 & 100 & 100 & 73 & 72.1 & 76.5 & 76.2   \\
          & 0.09 & 100 & 100 & 100 & 84.7 & 84.6 & 86.4 & 85.6   \\
  \bottomrule
\end{tabular}}
    \label{table6}
\end{table}

\section{Real Data Analysis}
In this section, we apply our proposed method to a pricing problem in finance, specifically exploring whether the weekly excess return of any asset is zero on average. Denote $X_{tj}=R_{tj}-\mbox{rf}_{t}$ by the excess return of the $j$th asset at time $t$ for $t=1,\dots,n$ and $j=1,\dots,p$, where $R_{tj}$ is the return on asset $j$ during period $t$ and $\mbox{rf}_{t}$ is the risk-free return rate of all asset during period $t$. We study the zero-factor pricing model
\begin{eqnarray*}
\bX_t=\bmu+\bxi_t, \quad t=1,\dots,n,
\end{eqnarray*}
where $\bX_t=(X_{t1},\dots,X_{tp})^\top$, $\bmu=(\mu_1,\dots,\mu_p)^\top$, and $\bxi_t=(\xi_{t1},\dots,\xi_{tp})^\top$ is the zero-mean error vector. We consider the null hypothesis test $H_0:\bmu=\mathbf{0}$, which means that, the return rate of any asset $R_{tj}$ is equal to the risk-free return rate $\mbox{rf}_{t}$ on average.
We analyzed the weekly return rates of stocks in the S\&P 500 index from January 14, 2005, to November 24, 2023. The weekly data were calculated using the stock prices on Fridays. After deleting missing values, we obtained a dataset with $p=393$ securities across $n=888$ observations.

\begin{figure}
\centering
\subfloat[An example of an individual figure sub-caption.]{%
\resizebox*{7cm}{!}{\includegraphics{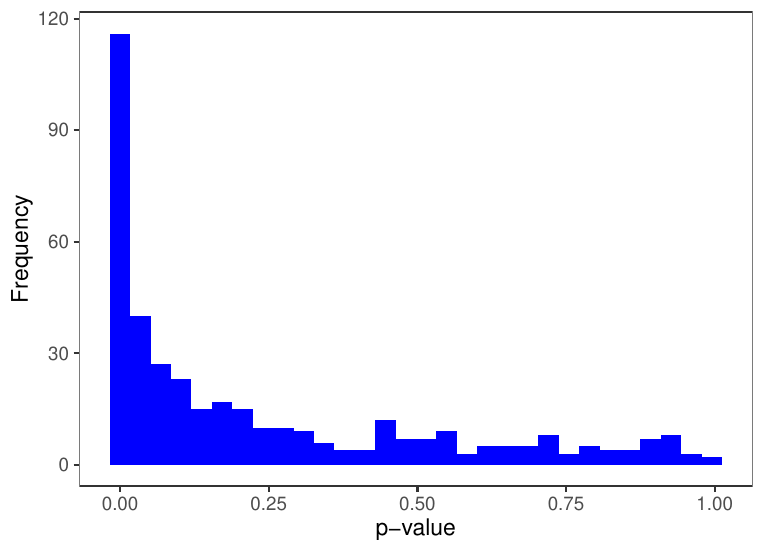}}}\hspace{5pt}
\subfloat[A slightly shorter sub-caption.]{%
\resizebox*{7cm}{!}{\includegraphics{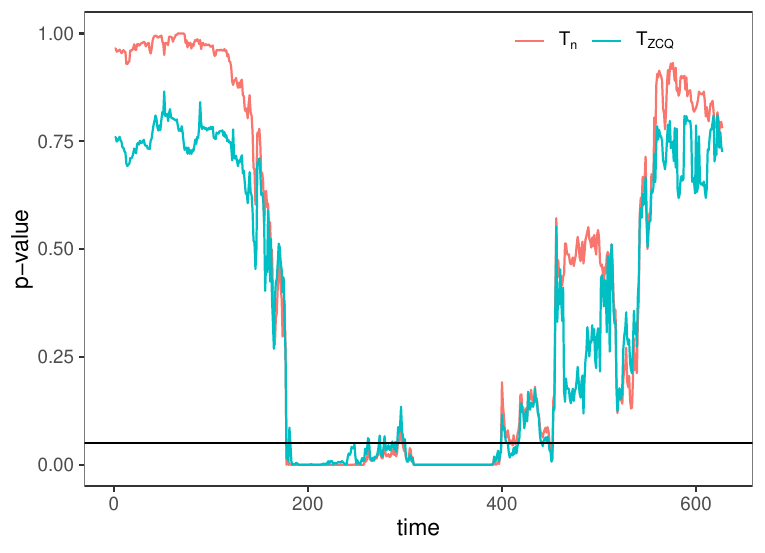}}}
\caption{U.S.'s weekly datasets.} \label{fig 1}
\end{figure}

First, we examine the presence of time series dependence in the observed sequences of our dataset. To achieve this, we apply the Box-Pierce test, a traditional method for detecting autocorrelation. The p-value histogram from the Box-Pierce test for the U.S. datasets is shown in Figure~\ref{fig 1}--(a). We observe a significant number of p-values below 0.05, indicating that some observed sequences may exhibit autocorrelation. Furthermore, we employ the max-type high-dimensional white noise test procedure, as described by \cite{feng2022testing}, to assess whether the excess return are white noise. This test produces a p-value of zero, strongly suggesting time dependence among the sequences. Consequently, test procedures that assume independent and identically distributed (i.i.d.) random vectors may not perform well and could yield inaccurate results.

We employed a rolling window procedure with a window length of $m=260$ weeks (approximately 5 years) to evaluate the performance of our proposed $T_n$ test and $T_{\mbox{\tiny{ZCQ}}}$ test. Figure~\ref{fig 1}--(b) displays the p-values for each test procedure over time, with the black horizontal line representing the 0.05 significance level.
In the first nine years and the last eight years, both $T_n$ test and $T_{\mbox{\tiny{ZCQ}}}$ test failed to reject the null hypothesis, suggesting that the weekly excess returns of any asset may be zero on average during these periods. However, during the ﬁve-year period in the middle, both the $T_n$ test and $T_{\mbox{\tiny{ZCQ}}}$ test frequently reject the null hypothesis, indicating that the weekly excess returns of any asset were not zero on average in the interval. It is noteworthy that our proposed test demonstrated a stronger rejection of the null hypothesis compared to the $T_{\mbox{\tiny{ZCQ}}}$ test, indicating higher sensitivity and potentially greater accuracy in detecting deviations from zero weekly excess returns.

\section{Conclusion}

In this study, we introduce a novel mean test method for high-dimensional data, accommodating both cross-sectional and time series dependencies. The asymptotic normality of the proposed test statistic is established as both $p$ and $n$ approach infinity concurrently. The tests presented in this paper follow a sum-type procedure, exhibiting robust performance under dense alternatives when the time series of a majority of variables are depart from zero. An intriguing challenge is to develop a max-type test procedure that performs well under sparse alternatives, where the time series of a handful of variables are depart from zero.

Moreover, several studies, including \cite{feng2024asymptotic} and \cite{chen2022asymptotic}, have explored the asymptotic independence between sum-type and max-type test procedures. Therefore, it would be fascinating to investigate the asymptotic independence between these two types of procedures in the context of high-dimensional unit root testing. %In addition, obtaining the asymptotic distribution of $T_{tr}$ under the alternative hypothesis poses a significant challenge. Specifically, when $\Pi$ is extremely small, the time series $x_t$ could take the form of $\Sigma^{1/2}Y_t$, making it closely resemble the null hypothesis. This issue warrants further investigation.

\appendix

\section{Notations}
%\subsection{Notations}
Let $\bepsilon_{t}=\bSig^{1/2}\bY_t$, we define a $M$-dependent approximation sequence for $\{\bepsilon_t\}_{t=1}^n$ as follows,
\[\bgamma_{t}^{(M)}=\E[\bepsilon_{t}|\bZ_{t-M},\dots,\bZ_{t}] = \bSig^{1/2}\sum_{k = 0}^{M} b_k\bZ_{t-k},\]
where $M = \lceil \min (n,p)^{1/8}\rceil $. In order to simplify notation, we omit the superscript in the absence of ambiguity. By replacing the $\bepsilon_t$ in $T_n$ with $\bgamma_t^{(M)}$, we define
$$
T_n^{(NG)} = \frac12(n\bar{\bgamma}_n^\top\bar{\bgamma}_n -\frac1n\sum_{t = 1}^n\bgamma_t^\top\bgamma_t),
$$
where $\bar{\bgamma}_n = n^{-1}\sum\nolimits_{t=1}^n \bgamma_t$.
Let $a_{h,M} = \sum_{k = 0}^{M-h} b_kb_{k+h}$, we define the auto-covariance matrices of $\bgamma_t^{(M)}$ at lag $h$ as
\[\bGam_{h,M}=\E[\bgamma_t\bgamma_{t+h}^\top] = a_{h,M}\bSig, \quad h = 0,1,\cdots,M,\]
and $\bOme_{n,M} = \bGam_{0,M} + 2\sum_{h = 1}^M (1-\frac hn) \bGam_{h,M}$. Let $\{\bdelta_t\}_{t=1}^n$ be a Gaussian sequence which is independent of $\{\bgamma_{t}\}_{t=1}^n$ and preserves the auto-covariance structure of $\{\bgamma_{t}\}_{t=1}^n$, i.e. $\E[\bdelta_t\bdelta_{t+h}^\top] = \bGam_{h,M}$. Similarly, with $\bar{\bdelta}_n = n^{-1}\sum\nolimits_{t=1}^n \bdelta_t$, we define
$$
T_n^{(G)} = \frac12(n\bar{\bdelta}_n^\top\bar{\bdelta}_n -\frac1n\sum_{t = 1}^n\bdelta_t^\top\bdelta_t).
$$

\section{Important Lemmas}
%\subsection{Important Lemmas}
We first establish the following lemmas which will be frequently used in our subsequent proofs.
\begin{Lem}\label{Lem EWBW2 leq E2WBW}
Let $\bW = \sum_{t = 0}^\infty a_t\bZ_{t}$, where $\sum_{t= 0}^\infty \Abs{a_t} < \infty$, there exists a positive constant $\tau_1\geq 3$ such that for all $p\times p$ positive semi-definite matrix $\bB$ satisfy
\begin{align*}
\E \Brab{(\bW^\top \bB \bW)^2} \leq \tau_1 \Brab{\E(\bW^\top \bB \bW)}^2.
\end{align*}
\end{Lem}

\begin{Prof}
Let $[B]_{ij}$ denotes the entry in row $i$, column $j$ of $\bB$. Since $\bB$ is positive semi-definite matrix, $\bB = \bD\bLam \bD^\top$, where $\bD$ is an orthogonal matrix and $\bLam = \diag(\lambda_1,\cdots,\lambda_p)$, $\lambda_1\geq \lambda_2\geq \cdots \geq \lambda_p\geq 0$.  Let $\bZ_t' = \bD^\top \bZ_t$, we can calculate their second and fourth moments
\[\E[{Z'_{t,j}}^2] = \E\Brab{\sum_{i = 1}^p[D]_{ij}Z_{t,i}}^2 = 1,\]
\[\E[{Z'_{t,j}}^4] = \E\Brab{\sum_{i = 1}^p[D]_{ij}Z_{t,i}}^4 = \sum_{i = 1}^p [D]_{ij}^4(\mu_4 - 3) + 3\leq \mu_4.\]
Then
\begin{align*}
   \E \Brab{(\bW^\top \bB \bW)} = \sum_{r = 1}^p [B]_{r,r} \E\Brab{\bigg(\sum_{i = 0}^\infty a_{i}Z_{i,r}\bigg)^2}
    = \sum_{r = 1}^p [B]_{r,r}\sum_{i = 0}^\infty a_{i}^2 = \sum_{i = 0}^\infty a_{i}^2 \tr(\bB),
\end{align*}
\begin{align*}
    \E \Brab{(\bW^\top \bB \bW)^2} = & \sum_{1\leq r_1,r_2,r_3,r_4\leq p} [B]_{r_1,r_2}[B]_{r_3,r_4}\E\Brab{\prod_{j = 1}^4\Braa{\sum_{i = 0}^\infty a_{i}Z_{i,r_j}}}\\
    = & \sum_{r = 1}^p [B]_{rr}^2 \E\Brab{\sum_{i = 0}^\infty a_{i}Z_{i,r}}^4
\end{align*}
\begin{align*}
    &+ \sum_{r\neq s}\Braa{2[B]_{r,s}^2 + [B]_{r,r}[B]_{s,s}}\E\Brab{\bigg(\sum_{i = 0}^\infty a_{i}Z_{i,r}\bigg)^2 \bigg(\sum_{i = 0}^\infty a_{i}Z_{i,s}\bigg)^2}\\
    = & \sum_{r = 1}^p [B]_{r,r}^2\bigg(\sum_{i = 0}^\infty a_{i}^4\mu_4 + \sum_{i\neq j}3a_{i}^2 a_j^2\bigg) + \sum_{r\neq s} \Braa{2[B]_{r,s}^2 + [B]_{r,r}[B]_{s,s}}\bigg(\sum_{i = 0}^\infty a_{i}^2\bigg)^2\\
    \leq& \tr(\bB)^2\bigg(\sum_{i = 0}^\infty a_{i}^4\mu_4 + \sum_{i\neq j}3a_{i}^2 a_j^2\bigg) + \Braa{2\tr(\bB^2) + \tr(\bB)^2}\bigg(\sum_{i = 0}^\infty a_{i}^2\bigg)^2\\
    \leq& (\mu_4 + 6)\bigg(\sum_{i = 0}^\infty a_{i}^2\bigg)^2\tr(\bB)^2 = (\mu_4 + 6)\Brab{\E (\bW^\top \bB\bW)}^2.
\end{align*}
This completes the proof by taking $\tau_1 = \mu_4+6$.
\end{Prof}

\begin{Lem}\label{Lem for fourth moment of eps}
Under Assumptions 1--3, for all $k_1,k_2,k_3,k_4$, there exists a constant $\tau_2$ such that
\[\Abs{\E\Brab{\bepsilon_{k_1}^\top\bepsilon_{k_2} - \tr\Braa{\bGam_{\Abs{k_1-k_2}}}}\Brab{\bepsilon_{k_3}^\top\bepsilon_{k_4} - \tr\Braa{\bGam_{\Abs{k_3-k_4}}}}} \leq \tau_2\tr\Braa{\bOme^2}.\]
\end{Lem}
\begin{Prof}
\begin{align*}
&\Abs{\E\Brab{\bepsilon_{k_1}^\top\bepsilon_{k_2} - \tr\Braa{\bGam_{\Abs{k_1-k_2}}}}\Brab{\bepsilon_{k_3}^\top\bepsilon_{k_4} - \tr\Braa{\bGam_{\Abs{k_3-k_4}}}}}\\
\leq &\bigg\{\E\Brab{\bepsilon_{k_1}^\top\bepsilon_{k_2} - \tr\Braa{\bGam_{\Abs{k_1-k_2}}}}^2\E\Brab{\bepsilon_{k_3}^\top\bepsilon_{k_4} - \tr\Braa{\bGam_{\Abs{k_3-k_4}}}}^2\bigg\}^{1/2}
\end{align*}
It suffice to prove, for all $h \geq 0$,
\[\E\Brab{\Braa{\bepsilon_{0}^\top\bepsilon_{h} - \tr\Braa{\bGam_{h}}}^2}\leq \tau_2\tr\Braa{\bOme^2}.\]
We split $\bepsilon_h$ into two independent parts
\[\bepsilon_h = \bSig^{1/2}\sum_{k = 0}^\infty b_{k+h}\bZ_{-k} + \bSig^{1/2}\sum_{k = 0}^{h-1}b_k\bZ_{h-k} := \bepsilon_{h,(1)} + \bepsilon_{h,(2)}.\]
Then we can simplify $\E\Brab{\Braa{\bepsilon_{0}^\top\bepsilon_{h} - \tr\Braa{\bGam_{h}}}^2}$ as follows,
\begin{align*}
\E\Brab{\Braa{\bepsilon_{0}^\top\bepsilon_{h} - \tr\Braa{\bGam_{h}}}^2} =\E \Brab{\Braa{\bepsilon_0^\top \bepsilon_{h,(1)} - \tr(\bGam_{h})}^2} +  \E\Brab{(\bepsilon_0^\top\bepsilon_{h,(2)})^2}.
\end{align*}
Since
\begin{align*}
\E\Brab{(\bZ_r^\top \bSig \bZ_s)^2} = &\tr(\bSig^2), \\
\E\Brab{(\bZ_k^\top \bSig \bZ_k  - \tr(\bSig))^2} = & 2\tr(\bSig^2) + (\mu_4 - 1)\tr(\bSig\odot\bSig) \leq (\mu_4 + 1)\tr(\bSig^2),
\end{align*}
where $\odot$ denotes Hadamard product.
Then, we have
\[\E\Brab{(\bepsilon_0^\top\bepsilon_{h,(2)})^2} = \Braa{\sum_{k = 0}^{h - 1} b_k^2}\Braa{\sum_{k = 0}^\infty b_k^2}\tr(\bSig^2)\]
and
\begin{align*}
&\E \Brab{\Braa{\bepsilon_0^\top \bepsilon_{h,(1)} - \tr(\bGam_{h})}^2} \\
= & \sum_{k = 0}^\infty b_k^2b_{k+h}^2\E[(\bZ_k^\top \bSig \bZ_k  - \tr(\bSig))^2] + \sum_{k,l=0, k\neq l}^\infty b_kb_{l+h}b_lb_{k+h}\E(\bZ_k^\top \bSig \bZ_l\bZ_l^\top \bSig \bZ_k) \\
& + \sum_{k,l=0, k\neq l}^\infty  b_k^2b_{l+h}^2\E(\bZ_k^\top \bSig \bZ_l\bZ_k^\top \bSig \bZ_l)  \\
\leq & (\mu_4 + 1)\sum_{k = 0}^\infty b_k^2b_{k+h}^2\tr(\bSig^2) + \sum_{k,l=0, k\neq l}^\infty(b_kb_{l+h}b_lb_{k+h}+b_k^2b_{l+h}^2)\tr(\bSig^2)\\
\leq & \bigg\{(\mu_4+1)\sum_{k = 0}^\infty b_k^2b_{k+h}^2 + \Braa{\sum_{r= 0}^\infty b_kb_{k+h}}^2 + \Braa{\sum_{k = 0}^\infty b_k^2}\Braa{\sum_{l = 0}^\infty b_{l+ h}^2}\bigg\}\tr(\bSig^2)\\
\leq & (\mu_4 + 3)\bigg(\sum_{k = 0}^\infty b_k^2\bigg)^2\tr(\bSig^2).
\end{align*}
It completes the proof with $\tr\Braa{\bOme^2}= \bigg(\sum\limits_{h_1 = -\infty}^{\infty}\sum\limits_{k = 0}^\infty b_kb_{k+h_1}\bigg)^2\tr(\bSig^2)$ and $b_n = o(n^{-5})$.
\end{Prof}

Let $b_{k,M} := b_{k}\mathbf{1}_{\{k\leq M\}}$, applying the similar technique analysis, we get the following lemma of $\bgamma_{k}$.

\begin{Lem}\label{Coro for fourth moment of gamma}
    Under Assumptions 1--3, for all $k_1,k_2,k_3,k_4$, there exists a constant $\tau_2$ such that
    \[\Abs{\E\Brab{\bgamma_{k_1}^\top\bgamma_{k_2} - \tr\Braa{\bGam_{\Abs{k_1-k_2},M}}}\Brab{\bgamma_{k_3}^\top\bgamma_{k_4} - \tr\Braa{\bGam_{\Abs{k_3-k_4},M}}}} \leq \tau_2\tr\Braa{\bOme_{n,M}^2}.\]
\end{Lem}
\begin{Lem}\label{Lem for the order of bn}
    Let $\{b_n\}_{n=0}^{\infty}$ is a sequence of numbers, if there exist a constant $k > 1$ such that $b_n = o(n^{-k})$, then
    \[\sum_{m = n}^\infty b_m = o(n^{-(k-1)}).\]
\end{Lem}
\begin{Prof}
For any $\epsilon>0$, there exists $N\in \N$, such that for all $n \geq N$, $n^{k}|b_n| \leq \epsilon$. Then for all $n \geq N$, $k >1$, we have
    %\begin{align*}
        %n^{k-1} \bigg|\sum_{m = n}^\infty b_m\bigg| = \sum_{m = 1}^\infty \frac1n \sum_{t = mn}^{(m+1)n-1} n^{k}b_t\leq \epsilon\sum_{m = 1}^\infty \frac1{m^k}  < \infty.
    %\end{align*}
    \begin{align*}
        n^{k-1} \bigg|\sum_{m = n}^\infty b_m\bigg| = \bigg|\sum_{m = 1}^\infty \frac1n \sum_{t = mn}^{(m+1)n-1} \frac{(mn)^k}{m^k}b_t\bigg|\leq \sum_{m = 1}^\infty \frac1n \sum_{t = mn}^{(m+1)n-1} \bigg|\frac{t^k}{m^k}b_t\bigg| \leq\epsilon\sum_{m = 1}^\infty \frac1{m^k}  < \infty.
    \end{align*}
\end{Prof}

\subsection{Proof of Theorem \ref{Th 1}}
Note that
\[\frac{T_n - \mu_n}{\sqrt{\sigma_n^2}} = \frac{\sqrt{\tr(\bOme_{n,M}^2)/2}}{\sqrt{\sigma_n^2}}\Braa{\frac{T_n^{(G)} - \mu_n}{\sqrt{\tr(\bOme_{n,M}^2)/2}} + \frac{T_n^{(NG)} - T_n^{(G)}}{\sqrt{\tr(\bOme_{n,M}^2)/2}} + \frac{T_n - T_n^{(NG)}}{\sqrt{\tr(\bOme_{n,M}^2)/2}} }.\]
Our proof is mainly divided into the following three parts: first, we prove the asymptotic normality of $T_n^{(G)}$ in Lemma \ref{Lem in part1}, then we use some Gaussian approximation approach to prove the asymptotic normality of $T_n^{(NG)}$ in Lemma \ref{Lem in part2}. Finally, we use $T_n^{(NG)}$ to approximate $T_n$ in Lemma \ref{Lem in part3}. Besides that, we need to prove $\tr(\bOme_{n,M}^2)  = \sigma_n^2(1+ o(1))$. Then, it completes the proof of Theorem \ref{Th 1}.

\subsubsection{Part 1}
\begin{Lem}\label{Lem in part1}
Under Assumptions 1--3 and $H_0$ holds, as $n \to \infty$,
\begin{equation}\label{eq for Ass G conclusion}
\frac{T_n^{(G)}- \E[T_n^{(G)}]}{\sqrt{\tr(\bOme_{n,M}^2)/2}} \limd N(0,1)
\end{equation}
\end{Lem}

\begin{Prof}
Based on simple calculations, we have $\E[T_n^{(G)}]=[\tr(\bOme_{n,M})-\tr(\bGam_{0,M})]/2$. Let $\tilde{\bdelta} = (\bdelta_1^\top ,\dots ,\bdelta_n^\top)^\top$, then $\tilde{\bdelta}\sim N(\mathbf{0}_{np}, \tilde{\bSig})$, where $\tilde{\bSig} = \sum_{h \in \mathcal{M}} \bD_h\otimes \bGam_{h,M}$, $\bD_h$ is the $n\times n$ matrix with $[D_h]_{ij} = 1$ if $i - j = h$ and $[D_h]_{ij} = 0$ otherwise. Let $\mathbf{1}_n$ be the $n$-dimensional vector with all elements one, $\bI_n$ be the $n\times n$ identity matrix, and $\mathcal{M}=\{-M,\dots,M\}$, we have
\begin{align*}
\Var(T_n^{(G)}) =& \frac{1}{4n^{2}}\Var\bigg(\tilde{\bdelta}^\top [\mathbf{1}_{n}\mathbf{1}_{n}^\top \otimes \bI_{p} - \bI_{np}]\tilde{\bdelta}\bigg) \\
=& \frac{1}{2n^{2}}\tr\bigg([\mathbf{1}_{n}\mathbf{1}_{n}^\top \otimes \bI_{p} - \bI_{np}]\tilde{\bSig}[\mathbf{1}_{n}\mathbf{1}_{n}^\top \otimes \bI_{p} - \bI_{np}]\tilde{\bSig}\bigg) \\
=& \frac{1}{2n^{2}}\sum_{a \in \mathcal{M}}\sum_{b \in \mathcal{M}}[\tr(\mathbf{1}_{n}^\top\bD_a\mathbf{1}_{n}\mathbf{1}_{n}^\top\bD_b\mathbf{1}_{n})-2\tr(\mathbf{1}_{n}^\top\bD_a\bD_b\mathbf{1}_{n})+\tr(\bD_a\bD_b)]\tr(\bGam_{a,M} \bGam_{b,M}) \\
=& \frac1{2} \sum_{a \in \mathcal{M}}\sum_{b \in \mathcal{M}}\Big[\Big(1-\frac{|a|}{n}\Big)\Big(1-\frac{|b|}{n}\Big) - \frac{2}{n}\Big(1-\frac{|a|}{n}\Big) \land \Big(1-\frac{|b|}{n}\Big)\\
& + \frac{1}{n}\Big(1-\frac{|a|}{n}\Big)\mathbf{1}_{\{a = b\}}\Big]\tr(\bGam_{a,M} \bGam_{b,M}).
\end{align*}
Therefore,
\begin{equation}\label{eq for Ass G var(T)}
\Var(T_N^{(G)}) = \frac{1}2 \tr(\bOme_{n,M}^2) (1 + o(1)),
\end{equation}
where $\bOme_{n,M} = \bGam_{0,M} + 2\sum_{h = 1}^M (1-\frac hn) \bGam_{h,M}$.
Now, it is suffices to prove
\begin{equation}\label{Lem1-E1}
 \frac{n\bar{\bdelta}_n^\top\bar{\bdelta}_n - \tr(\bOme_{n,M})}{\sqrt{2\tr(\bOme_{n,M}^2)}}  \limd N(0,1), \quad\quad
 \frac{\frac1n\sum_{t = 1}^n\bdelta_t^\top\bdelta_t - \tr(\bGam_{0,M})}{\sqrt{2\tr(\bOme_{n,M}^2)}} = o_p(1).
\end{equation}
We will apply Theorem 3.1 in \cite{ayyala2017mean} and its correction Theorem 2.1 in \cite{cho2019note} to prove the first part in equation (\ref{Lem1-E1}). Before that, we rephrase Theorem 2.1 in \cite{cho2019note} with the notation in our paper as the following Proposition \ref{prop 1}.
\begin{Prop}[\citealt{cho2019note}, Theorem 2.1]
\label{prop 1}
Let $\bdelta_1,\dots,\bdelta_n\in\mathbb{R}^p$ follow an $M$-dependent strictly stationary Gaussian process with zero mean and autocovariance structure given by $\bGam_{0,M},\dots,\bGam_{M,M}$. Assume that $p = O(n)$ and $p \to \infty$, $M = O(n^{1/8})$ and
\begin{equation}\label{eq for Ayyla Ga4AndOmega}
\tr(\bGam_{a,M}\bGam_{b,M}\bGam_{c,M}\bGam_{d,M}) = o(M^{-4}\tr^2(\bOme_{n,M}^2))
\end{equation}
then as $n\to \infty$,
\begin{equation*}\label{eq for ayyla eq10 modifty}
\frac{n\bar{\bdelta}_n^\top\bar{\bdelta}_n - \tr(\bOme_{n,M})}{\sqrt{2\tr(\bOme_{n,M}^2)}}  \limd N(0,1).
\end{equation*}
\end{Prop}
%\begin{Rmk}
    %The above proportion is obtained by making some minor modification to formula (10) in the proof of theorem 2.1 in \cite{cho2019note}. This equation has the following original form
    %\begin{equation}\label{eq for ayyla eq10}
            %\frac{\bar{\epsilon}_n^\top\bar{\epsilon}_n - \frac1n \tr(\Omega_{n,M})}{\sqrt{\var(\mathcal{M}_n)}}  \limd N(0,1)
    %\end{equation}
%Actually, we don't need to know what the $\mathcal{M}_n$ means. Proposition 3.1 in \cite{ayyala2017mean} shows $\var(\mathcal{M}_n) = \frac2n\tr(\Omega_{n,M}^2)(1+o(1))$, take it into \eqref{eq for ayyla eq10}, we get \eqref{eq for ayyla eq10 modifty}.
%\end{Rmk}
In order to utilize the above Proposition~\ref{prop 1}, we only need to prove that the condition~\eqref{eq for Ayyla Ga4AndOmega} holds. It suffices to prove that
\[\tr(\bSig^4) = o(M^{-4}\tr^2(\bSig^2)).\]
Let $\lambda_1\geq\cdots\geq\lambda_p$ be the eigenvalues of $\bSig$, by Assumption 2, we have $\lambda_1 \leq C_0$, and $\tr(\bSig)\geq p C_1$. Since $ \tr^2(\bSig) \leq p\tr(\bSig^2)$, $M = \lceil \min (n,p)^{1/8}\rceil$, we have
\[\frac{M^{4}\tr(\bSig^4)}{\tr^2(\bSig^2)} \leq \frac{p^{1/2}\sum_{i = 1}^p\lambda_i^4}{p^{-2}\tr^4(\bSig)}\leq \frac{p^{3/2}C_0^4}{p^2C_1^4} = o(1).\]
Thus, the first part in equation \eqref{Lem1-E1} has been proved. Now, we prove the second part in equation \eqref{Lem1-E1}.
\begin{align*}
&\E\Brab{\bigg(\frac1n \sum_{t = 1}^n\bdelta_t^\top\bdelta_t - \tr(\bGam_{0, M})\bigg)^2}\\%= \frac 1{n^2}\Var\Braa{\tilde{\bdelta}^\top \tilde{\bdelta}}\\
%=& \frac 2{n^2}\sum_{a \in \mathcal{M}}\sum_{b \in \mathcal{M}} \tr((D_aD_b) \otimes (\Gamma_{a,M}\Gamma_{b,M})) \\
=& \frac 2{n^2}\sum_{a \in \mathcal{M}}\sum_{b \in \mathcal{M}} \tr(\bD_a \bD_b) \tr( \bGam_{a,M} \bGam_{b,M})\\
= & \frac 2{n}\sum_{ h \in \mathcal{M}}\Big(1-\frac{h}{n}\Big)\tr(\bGam_{h,M}^2) = o(\tr(\bOme_{n,M}^2)).
\end{align*}
This completes the proof.
\end{Prof}

\begin{Lem}
Under Assumptions 1--3 and $H_0$ holds, as $\min(n,p) \to \infty$,
\begin{equation}
\E[T_n^{(G)}] - \mu_n = o(\tr^{1/2}(\bOme_{n,M}^2)).
\end{equation}
\end{Lem}
\begin{Prof}
By Lemma \ref{Lem for the order of bn}, we have $\sum\nolimits_{i=M}^{\infty} |b_i| = o(M^{-4})$, then, as $\min(n,p) \to \infty$,
\begin{align*}
\Abs{\E[T_n^{(G)}] - \mu_n} = & \Abs{\sum_{h = 1}^M \frac{n-h}n \tr(\bGam_h - \bGam_{h,M})} = \tr(\bSig)\Abs{\sum_{h = 1}^M \frac{n-h}n\sum_{i = M+1}^\infty b_ib_{i-h}}\\
\leq & p^{1/2}\tr^{1/2}(\bSig^2) \bigg(\sum_{i = M+1-h}^\infty \Abs{b_i}\bigg)\bigg(\sum_{i = 1}^\infty \Abs{b_i}\bigg)\\
= & o\{p^{1/2}M^{-4}\tr^{1/2}(\bSig^2)\} = o(\tr^{1/2}(\bOme_{n,M}^2)).
\end{align*}
\end{Prof}

\subsubsection{Part 2}
In this part, we will use the Gaussian approximation technique to prove the asymptotic normality of $T_n^{(NG)}$, which can be stated as the following Lemma \ref{Lem in part2}.
\begin{Lem}\label{Lem in part2}
Under Assumptions 1--3 and $H_0$ holds, as $n \to \infty$,
\[T_n^{(G)} - T_n^{(NG)} = o_p(\tr^{1/2}(\bOme_{n,M}^2)).\]
\end{Lem}
%When our sample is independent and equally distributed (i.e. $M = 0$), \cite{wang2022approximate} presents a method for approximating two half-quadratic forms composed of Gaussian and non-Gaussian samples. We modify the proof of this paper, and prove the $M$-independent situation. Before proving this lemma, we first define some special structures about the components of $T$.
%In this part, we only discuss the related properties of $\{\bdelta_t\}$ and $\{\bgamma_t\}$, and not deal with $\{\bepsilon\}$.
\begin{Prof}
Since neither $\bGam_h$ nor $\bOme_{n}$ appears in this part, we omit the subscript $M$ from $\bGam_{h,M}$ and abbreviate $\bOme_{n,M}$ as $\bOme_n$ for simplicity. Define
\[\mathcal{F}_{ij}^{(G)} = n^{-1} \Brab{\bdelta_i^\top\bdelta_j - \tr(\bGam_{\Abs{i-j}})}.\]
For any $n$, choose $\alpha\in (0,1)$ and $C > 0$ and $\omega_n = Cn^\alpha > M$, so that $n = \omega_nq_n + r_n$. Then we can define the following random variables,
\begin{align*}
    B_{ij}^{(G)} & = \sum_{k = (i-1)\omega_n+1}^{i\omega_n-M}\sum_{l = (j-1)\omega_n +1}^{j\omega_n-M} \mathcal{F}_{kl}^{(G)}, \\
    D_{ij}^{(G)} & = \sum_{k = (i-1)\omega_n+1}^{i\omega_n}\sum_{l = (j-1)\omega_n +1}^{j\omega_n} \mathcal{F}_{kl}^{(G)} - B_{ij}^{(G)}, \\
    F^{(G)} & = \sum_{(i,j) \in \{1,\cdots,n\}^2\backslash \{1,\cdots,\omega_n q_n\}^2 }\mathcal{F}_{ij}^{(G)}.
\end{align*}
For non-Gaussian sequence $\{\bgamma_t\}_{t=1}^n$, we define $B_{ij}^{(NG)},D_{ij}^{(NG)},F^{(NG)}$, similarly. For the sake of simplicity, when we don't emphasize the difference between Gaussian and non-Gaussian processes, we use the unmarked symbols $B_{ij}$ to represent $B_{ij}^{(G)}$ or $B_{ij}^{(NG)}$. Let
\[S_1 = \sum_{1\leq i < j \leq q_n}B_{ij}, \quad S_2 = \sum_{i = 1}^{q_n} \Braa{B_{ii} - \sum_{k = (i-1)\omega_n+1}^{i\omega_n-M}\mathcal{F}_{kk}} \]  \[ S_3 = \sum_{1\leq i,j \leq q_n} D_{ij} - \sum_{i = 1}^{q_n}\sum_{k = i\omega_n -M+1}^{i\omega_n}\mathcal{F}_{kk}, \quad S_4 = F - \sum_{k = q_n\omega_n+1}^n\mathcal{F}_{kk},\]
and $\Delta S_i = S_i^{(G)} - S_i^{(NG)}$. Then, $T_n^{(G)} - T_n^{(NG)} = \Delta S_1 + \Delta S_2/2 + \Delta S_3/2 + \Delta S_4/2$. Now, it suffice to prove that $\Delta S_i = o_p(\tr^{1/2}(\bOme_{n,M}^2))$, which is presented in Lemmas \ref{Lem for NG to G 1}--\ref{Lem for NG to G 4}.
\end{Prof}

\begin{Lem}\label{Lem for NG to G 1}
Under Assumptions 1--3 and $H_0$, as $n\to\infty$,
\begin{equation}\label{eq for Lem for NG to G 1}
\Delta S_1 = o_p(\tr^{1/2}(\bOme_{n,M}^2)).
\end{equation}
\end{Lem}
\begin{Prof}
For $i = 1, \dots, q_n$, we define
\begin{eqnarray*}
\bbeta_{i} = \frac{1}{\omega_n - M}\sum_{k = (i-1)\omega_n +1}^{i\omega_n-M}\bdelta_k, \quad\quad \bxi_{i} = \frac{1}{\omega_n - M}\sum_{k = (i-1)\omega_n +1}^{i\omega_n-M}\bgamma_k.
\end{eqnarray*}
Since $\{\bbeta_i\}_{i=1}^{q_n}$ and $\{\bxi_i\}_{i=1}^{q_n}$ are both independent series, for any  $i\neq j$,
\begin{eqnarray*}
B_{ij}^{(G)}  = \frac{(\omega_n - M)^{2}}{n}\bbeta_i^\top\bbeta_j, \quad  B_{ij}^{(NG)}  = \frac{(\omega_n - M)^{2}}{n}\bxi_i^\top\bxi_j.
\end{eqnarray*}
Define $\nu_n(\bx,\by) = \frac{(\omega_n-M)^2}{n\tr^{1/2}(\bOme_{n,M}^2)}\bx^\top \by$, and
\begin{eqnarray*}
W(\bbeta_1,\cdots,\bbeta_{q_n}) = \sum_{1\leq i<j\leq q_n} B_{ij}^{(G)}/\tr^{1/2}(\bOme_{n,M}^2) =  \sum_{1\leq i<j\leq q_n} \nu_n(\bbeta_i,\bbeta_j),\\
W(\bxi_1,\cdots,\bxi_{q_n}) = \sum_{1\leq i<j\leq q_n} B_{ij}^{(NG)}/\tr^{1/2}(\bOme_{n,M}^2)=  \sum_{1\leq i<j\leq q_n} \nu_n(\bxi_i,\bxi_j).
\end{eqnarray*}
By \cite{Pollard1984ConvergenceOS}, It is known that a sequence of random variables $\{Z_n\}_{n=1}^{\infty}$ converges weekly to a random variable $Z$ if and only if for every $f\in\mathcal{C}_b^3(\mathbb{R})$, $\E(f(Z_n))\rightarrow \E(f(Z))$. Thus, it is necessary to prove $\Abs{\E [f(W(\bxi_1,\cdots,\bxi_{q_n}))] - \E [f(W(\bbeta_1,\cdots,\bbeta_{q_n})) ]}\rightarrow 0.$
Since both $\bbeta_i$ and $\bxi_i$ are finite linear combination of $Z_{i,j}$, it's easy to check that $\{\bbeta_i\}_{i=1}^{q_n}$ and $\{\bxi_i\}_{i=1}^{q_n}$ satisfy the following Proportions \ref{Prop for NG to G upper bound of second moment}--\ref{Prop for existence of rho}.
For $k = 1, \cdots, q_n+1$, we further define
\begin{eqnarray*}
W_k &=& W(\bbeta_1,\cdots,\bbeta_{k-1},\bxi_k,\cdots,\bxi_{q_n}),\\
W_{k,0} &=& \sum_{1\leq i<j\leq k - 1}\nu_n(\bbeta_i,\bbeta_j) + \sum_{k+1\leq i<j\leq q_n}\nu_n(\bxi_i,\bxi_j) + \sum_{1\leq i\leq k-1}\sum_{k+1\leq j\leq q_n}\nu_n(\bbeta_i,\bxi_j).
\end{eqnarray*}
For a fixed function $f\in \mathcal{C}_b^3(\mathbb{R})$, we have
\[\Abs{\E [f(W(\bxi_1,\cdots,\bxi_{q_n}))] - \E [f(W(\bbeta_1,\cdots,\bbeta_{q_n})) ]}\leq \sum_{k = 1}^{q_n}\Abs{\E f(W_k) - \E f(W_{k+1})}.\]
By Taylor's expansion,
\[\Abs{ f(W_k) - f(W_{k,0}) - \sum_{i = 1}^2 \frac1{i!}\Braa{ W_k - W_{k,0}}^i f^{(i)}(W_{k,0}) }\leq \frac{\sup_{x\in \mathbb{R}}\Abs{f^{(3)}(x)}}6 \Abs{W_k - W_{k,0}}^3,\]
\[\Abs{ f(W_{k+1}) - f(W_{k,0}) - \sum_{i = 1}^2 \frac1{i!}\Braa{ W_{k+1} - W_{k,0}}^i f^{(i)}(W_{k,0}) }\leq \frac{\sup_{x\in \mathbb{R}}\Abs{f^{(3)}(x)}}6 \Abs{W_{k+1} - W_{k,0}}^3.\]
We have
\[\E[W_k - W_{k,0}|\bbeta_1,\cdots,\bbeta_{k-1},\bxi_{k+1},\cdots,\bxi_{q_n}]=\sum_{i = 1}^{k-1}\E[\nu_{n}(\bbeta_i,\bxi_k)|\bbeta_i] + \sum_{j = k+1}^{q_n} \E[\nu_{n}(\bxi_k,\bxi_j)|\bxi_j]=0.\]
Similarly, we have \[\E[W_{k+1} - W_{k,0}|\bbeta_1,\cdots,\bbeta_{k-1},\bxi_{k+1},\cdots,\bxi_{q_n}] = 0.\]
Using the similar method, we can prove that
\[\E[(W_k - W_{k,0})^2|\bbeta_1,\cdots,\bbeta_{k-1},\bxi_{k+1},\cdots,\bxi_{q_n}] = \E[(W_{k+1} - W_{k,0})^2|\bbeta_1,\cdots,\bbeta_{k-1},\bxi_{k+1},\cdots,\bxi_{q_n}].\]
Thus,
\[\E\Brab{\sum_{i = 1}^2 \frac1{i!} (W_k - W_{k,0})^if^{(i)}(W_{k,0}) } = \E\Brab{\sum_{i = 1}^2 \frac1{i!} (W_{k+1} - W_{k,0})^if^{(i)}(W_{k,0}) }.\]
Then we have
\begin{align*}
&\Abs{\E f(W(\bxi_1,\cdots,\bxi_{q_n})) - \E f(W(\bbeta_1,\cdots,\bbeta_{q_n})) }\leq\sum_{k = 1}^{q_n}\Abs{\E f(W_k) - \E f(W_{k+1})}\\
\leq &\sum_{k = 1}^{q_n}\frac{\sup_{x\in\R}\Abs{f^{(3)}(x)}}6\Braa{\E\Abs{W_k - W_{k,0}}^3 + \E\Abs{W_{k+1} - W_{k,0}}^3}\\
\leq &\sum_{k = 1}^{q_n}\frac{\sup_{x\in\R}\Abs{f^{(3)}(x)}}6\Brac{\Brab{\E\Abs{W_k - W_{k,0}}^4}^{3/4} +\Brab{\E\Abs{W_{k+1} - W_{k,0}}^4}^{3/4}}.
\end{align*}
Now we derive upper bounds for $\E\Abs{W_k - W_{k,0}}^4$ and $\E\Abs{W_{k+1} - W_{k,0}}^4$, by Proposition \ref{Prop for NG to G upper bound of second moment}, we have
\begin{align*}
&\E\Abs{W_k - W_{k,0}}^4\\
= & \sum_{i = 1}^{k-1}\E\Brab{\nu_n(\bbeta_i,\bxi_k)^4} + \sum_{j = k+1}^{q_n} \E\Brab{\nu_n(\bxi_k,\bxi_j)^4}+ 6\sum_{i_1 = 1}^{k-1}\sum_{i_2 = i_1+1}^{k-1}\E\Brab{\nu_n(\bbeta_{i_1},\bxi_k)^2\nu_n(\bbeta_{i_2},\bxi_k)^2}\\
& + 6\sum_{j_1 = k+1}^{q_n}\sum_{j_2 = j_1+1}^{q_n}\E\Brab{\nu_n(\bxi_k,\bxi_{j_1})^2\nu_n(\bxi_k,\bxi_{j_2})^2} + 6\sum_{i = 1}^{k-1}\sum_{j = k+1}^{q_n}\E\Brab{\nu_n(\bbeta_{i},\bxi_k)^2\nu_n(\bxi_k,\bxi_j)^2}\\
\leq & \rho \Bigg(\sum_{i = 1}^{k-1}\phi_{i,k}^4 + \sum_{j = k+1}^{q_n}\phi_{k,j}^4 + 6\sum_{i_1 = 1}^{k-1}\sum_{i_2 = i_1 + 1}^{k-1}\phi_{i_1,k}^2\phi_{i_2,k}^2
+ 6\sum_{j_1 = k+1}^{q_n}\sum_{j_2 = j_1+1}^{q_n}\phi_{k,j_1}^2\phi_{k,j_2}^2+ 6\sum_{i = 1}^{k-1}\sum_{j = k+1}^{q_n}\phi_{i,k}^2\phi_{k,j}^2\Bigg)\\
\leq & 3\rho\Braa{\sum_{i = 1}^{k-1}\phi_{i,k}^2 + \sum_{j = k+1}^{q_n}\phi_{k,j}^2}^2 = 3\rho\Inf_k^2,
\end{align*}
where $\Inf_k= \sum_{i = 1}^{k-1}\phi_{i,k}^2 + \sum_{j = k+1}^{q_n}\phi_{k,j}^2$. Similarly, we have $\E\Abs{W_{k+1} - W_{k,0}}^4 \leq 3\rho \Inf_k^2$. Thus,
\[\Abs{\E f(W(\bxi_1,\cdots,\bxi_{q_n})) - \E f(W(\bbeta_1,\cdots,\bbeta_{q_n})) }=O\bigg(\sum_{k = 1}^{q_n}\Inf_k^{3/2}\bigg).\]
Since $\bxi_i$ has the same first two moment as $\bbeta_i$, we have
\begin{eqnarray*}
\phi_{i, j}^2 &=& \E\Brab{\nu_n(\bxi_i,\bxi_j)^2} = \frac{(\omega_n - M)^4}{n^2\tr(\bOme_{n,M}^2)}\E\Brab{\bxi_i^\top \bxi_j\bxi_i^\top \bxi_j} \\
&=& \frac1{n^2\tr(\bOme_n^2)}\E\Brab{\sum_{k_1 = (i-1)\omega_n +1}^{i\omega_n-M}\sum_{l_1 = (j-1)\omega_n +1}^{j\omega_n-M}\sum_{k_2 = (i-1)\omega_n +1}^{i\omega_n-M}\sum_{l_2 = (j-1)\omega_n +1}^{j\omega_n-M}\bgamma_{k_1}^\top \bgamma_{l_1}\bgamma_{k_2}^\top \bgamma_{l_2}}\\
&=& \frac1{n^2\tr(\bOme_n^2)}\tr\Brab{\Braa{\sum_{k_1 = (i-1)\omega_n +1}^{i\omega_n-M}\sum_{k_2 = (i-1)\omega_n +1}^{i\omega_n-M} \bGam_{k_1 - k_2}}^2} \\
&=& \frac{(\omega_n - M)^2}{n^2}\cdot\frac{\tr(\bOme_{\omega_n-M,M}^2)}{\tr(\bOme_{n,M}^2)}\{1+o(1)\}.
\end{eqnarray*}
Then we have
\[\sum_{k = 1}^{q_n}\Inf_k^{3/2} = O\bigg[q_n\bigg\{\frac{(q_n-1)(\omega_n - M)^2\tr(\bOme_{\omega_n-M,M}^2)}{n^2\tr(\bOme_{n,M}^2)}\bigg\}^{3/2}\bigg] = o(1).\]
This completes the proof of Lemma \ref{Lem for NG to G 1}.
\end{Prof}

\begin{Prop}\label{Prop for NG to G upper bound of second moment}
For $1\leq i \leq q_n$,
    \begin{itemize}
        %\item $\E[\nu_n(\bxi_i,\bxi_j)^4] <\infty$
        \item For any $\ba\in \R^p$, $\E[\nu_n(\bxi_i, \ba)] = \E[\nu_n(\ba,\bxi_i)] = \E[\nu_n(\bbeta_i,\ba)] = \E[\nu_n(\ba,\bbeta_i)] = 0.$
         \item For any $\ba, \bb\in \R^p$,
    \begin{align*}
        &\E[\nu_n(\ba,\bxi_i)\nu_n(\bb,\bxi_i)] = \E[\nu_n(\ba,\bbeta_i)\nu_n(\bb,\bbeta_i)],\\
        &\E[\nu_n(\ba,\bxi_i)\nu_n(\bxi_i,\bb)] = \E[\nu_n(\ba,\bbeta_i)\nu_n(\bbeta_i,\bb)],\\
        &\E[\nu_n(\bxi_i,\ba)\nu_n(\bxi_i,\bb)] = \E[\nu_n(\bbeta_i,\ba)\nu_n(\bbeta_i,\bb)].
    \end{align*}
    \end{itemize}
\end{Prop}
%\begin{Prop}\label{Prop for NG to G upper bound of fourth moment}
%For $1\leq i<j\leq q_n$,
    %\begin{itemize}
    %\item $\E[\nu_n(\bxi_i,\bbeta_j)^4] < \infty,\E[\nu_n(\bbeta_i,\bxi_j)^4] <\infty,\E[\nu_n(\bbeta_i,\bbeta_j)^4] < \infty.$
    %\item For and $\ba\in\R^p,\E[\nu_n(\bbeta_i,\ba)] = \E[\nu_n(\ba,\bbeta_j)] = 0.$
   % \item for any $\ba, \bb\in \R^p$,
    %\begin{align*}
        %&\E[\nu_n(\ba,\bxi_k)\nu_n(\bb,\bxi_k)] = \E[\nu_n(\ba,\bbeta_k)\nu_n(\bb,\bbeta_k)],\quad\forall 1\leq k\leq q_n,\\
        %&\E[\nu_n(\ba,\bxi_j)\nu_n(\bxi_j,\bb)] = \E[\nu_n(\ba,\bbeta_j)\nu_n(\bbeta_j,\bb)],\quad\forall 1\leq j\leq q_n,\\
        %&\E[\nu_n(\bxi_i,\ba)\nu_n(\bxi_i,\bb)] = \E[\nu_n(\bbeta_i,\ba)\nu_n(\bbeta_i,\bb)],\quad \forall 1\leq i\leq q_n.
    %\end{align*}
%\end{itemize}
%\end{Prop}
\begin{Prop}\label{Prop for existence of rho}
    For $1\leq i <j\leq q_n$, let $\phi_{i,j}^2 = \E\Brab{\nu_n(\bxi_i,\bxi_j)^2}$, and $\rho_0 = \max\{\tau_1,3\}$. Under Assumptions 1--3,
    \[\max\Brac{\E[\nu_n(\bxi_i,\bxi_j)^4],\E[\nu_n(\bxi_i,\bbeta_j)^4],\E[\nu_n(\bbeta_i,\bxi_j)^4],\E[\nu_n(\bbeta_i,\bbeta_j)^4]} \leq \rho_0\phi_{i,j}^4\]
\end{Prop}
\begin{Prof}
Let $a_i$ be the coefficient of $\bZ_{t\omega_n - M -i}$ in $\bxi_t$. By Lemma \ref{Lem EWBW2 leq E2WBW}, for any semi-positive definite matrix $\bB$, we have
 \[\E[(\bxi_t^\top \bB \bxi_t)^2] \leq \tau_1 (\E[\bxi_t^\top \bB \bxi_t])^2,\]
 where $\tau$ is a constant.
It follows that,
 \begin{align*}
    \E\Brab{(\bxi_i^\top \bxi_j)^4} & = \E\Brab{\E[(\bxi_i^\top\bxi_j\bxi_j^\top\bxi_i)^2|\bxi_j]}\\
    &\leq \tau_1 \E\Brab{\E[(\bxi_i^\top\bxi_j\bxi_j^\top\bxi_i)|\bxi_j]^2} \\
    & = \tau_1 \E\Brab{(\bxi_j^\top \E[\bxi_i\bxi_i^\top ]\bxi_j)^2} = \tau_1\E^2[(\bxi_i^\top\bxi_j)^2].
\end{align*}
Since $\{\bbeta_i\}$ is i.i.d. Gaussian distributed random vector sequence, we have
\begin{align*}
    \E\Brab{(\bbeta_i^\top \bbeta_j)^4} & = \E\Brab{\E[(\bbeta_i^\top\bbeta_j\bbeta_j^\top\bbeta_i)^2|\bbeta_j]}\\
    &= \E\Brab{2\tr\Braa{(\E[\bbeta_i\bbeta_i^\top ]\bbeta_j\bbeta_j^\top )^2} + \Braa{\tr\Braa{\E[\bbeta_i\bbeta_i^\top ]\bbeta_j\bbeta_j^\top }}^2}\\
    &= 3\E\Brab{\Braa{\bbeta_j^\top \E[\bbeta_i\bbeta_i^\top ]\bbeta_j}^2} = 3\E[(\bbeta_i^\top\bbeta_j)^2]
\end{align*}
Similarly, we have $\E\Brab{(\bbeta_i^\top \bxi_j)^4} \leq \tau_1\E[(\bbeta_i^\top\bxi_j)^2]$ and $\E\Brab{(\bxi_i^\top \bbeta_j)^4} \leq \tau_1\E[(\bxi_i^\top\bbeta_j)^2]$, which completes the proof of Proportion \ref{Prop for existence of rho}.
\end{Prof}

\begin{Lem}\label{Lem for NG to G 2}
Under Assumptions 1--3 and $H_0$ holds, as $n \to \infty$,
\begin{equation}\label{eq for D_2 1}
\Delta S_2 = o_p(\tr^{1/2}(\bOme_{n,M}^2)).
\end{equation}
\end{Lem}
\begin{Prof}
We only need to prove
\begin{align}\label{eq for D_2 1}
&\displaystyle \sum_{i = 1}^{q_n}\sum_{(i-1)\omega_n + 1\leq k<l\leq i\omega_n - M}\mathcal{F}^{(G)}_{kl}= o_p(\tr^{1/2}(\bOme_{n,M}^2)),\nonumber\\
&\displaystyle \sum_{i = 1}^{q_n}\sum_{(i-1)\omega_n + 1\leq k<l\leq i\omega_n - M}\mathcal{F}^{(NG)}_{kl}= o_p(\tr^{1/2}(\bOme_{n,M}^2)).
\end{align}
Note that, $\{\sum_{(i-1)\omega_n + 1\leq k<l\leq i\omega_n - M}\mathcal{F}_{kl}\}_{i=1}^{q_n}$ are independent identically distributed sequence, and $\sum_{1\leq k<l\leq \omega_n - M}\mathcal{F}^{(G)}_{kl} = T^{(G)}_{\omega_n - M}-\E[T^{(G)}_{\omega_n - M}]$. Replace $n$ in { \eqref{eq for Ass G var(T)}} with $\omega_n - M$, we have
\[\Var\Braa{\sum_{i = 1}^{q_n}\sum_{(i-1)\omega_n + 1\leq k<l\leq i\omega_n - M}[\bdelta_k^\top\bdelta_l-\tr(\bGam_{k-l,M})]} = \frac{q_n(\omega_n-M)^2}{2}\tr\Braa{\bOme_{\omega_n-M,M}^2}(1+o(1)).\]
Then
\[\frac{q_n(\omega_n-M)\tr\Braa{\bOme_{\omega_n-M,M}^2}(1+o(1))}{2n^2\tr\Braa{\bOme_{n,M}^2}} = o(1),\]
which implies the first equation of \eqref{eq for D_2 1}. Now, we consider the second equation of \eqref{eq for D_2 1}.
Let $k_{(1)}<\cdots<k_{(4)}$ be the sorted value of $k_1,k_2,k_3,k_4$, and $d_i = k_{(i+1)} - k_{(i)}$.
Define $\mathcal{A}=\{(k_1,k_2,k_3,k_4): 1\leq k_1\neq k_2 \leq\omega_n - M; 1\leq k_3\neq k_4 \leq\omega_n - M\}$, $\mathcal{A}_1 =\Brac{d_1> M\text{ or } d_3> M}$, $\mathcal{A}_2 =\Brac{d_1\leq M, d_3\leq M, d_2> M}$, $\mathcal{A}_3 =\Brac{d_i\leq M , i = 1,2,3}$. We partition $\mathcal{A}_2$ into $\mathcal{A}_{2,1} = \mathcal{A}_2\cap\{\Abs{k_1 - k_2} \leq M, \Abs{k_3-k_4}\leq M\}$, and $\mathcal{A}_{2,2} =\mathcal{A}_{2}\setminus\mathcal{A}_{2,1}$.
For any $(k_1,\dots,k_4)\in\mathcal{A}_{1}$, $\E(\mathcal{F}_{k_1,k_2}\mathcal{F}_{k_3,k_4}) = 0$, and any $(k_1,\dots,k_4)\in\mathcal{A}_{2,1}$, $\E(\mathcal{F}_{k_1k_2}\mathcal{F}_{k_3k_4}) = \E(\mathcal{F}_{k_1k_2})\E(\mathcal{F}_{k_3k_4}) = 0.$ Then we only need to check $\mathcal{A}_{2,2}$ and $\mathcal{A}_{3}$. Since $|\mathcal{A}_{2,2}|\leq 10\omega_n^2M^2$ and $|\mathcal{A}_{3}|\leq 24\omega_n M^3$, we have
\begin{align*}
&\Var\Braa{\sum_{i = 1}^{q_n}\sum_{(i-1)\omega_n + 1\leq k<l\leq i\omega_n - M}\Brab{\bgamma_k^\top \bgamma_l - \tr(\bGam_{k-l,M})}}  \\
= & \frac{q_n}{4}\Bigg|\sum_{(k_1,\dots,k_4)\in\mathcal{A}} \E\big[\{\bgamma_{k_1}^\top \bgamma_{k_2} - \tr(\bGam_{k_1-k_2,M})\}\{\bgamma_{k_3}^\top \bgamma_{k_4} - \tr(\bGam_{k_3-k_4,M})\}\big]\Bigg|\\
= & \frac{n^2q_n}{4}\Bigg|\Bigg(\sum_{(k_1,\dots,k_4)\in\mathcal{A}\cap\mathcal{A}_1} + \sum_{(k_1,\dots,k_4)\in\mathcal{A}\cap\mathcal{A}_2} + \sum_{(k_1,\dots,k_4)\in\mathcal{A}\cap\mathcal{A}_3}\Bigg)\E(\mathcal{F}_{k_1 k_2}\mathcal{F}_{k_3 k_4})\Bigg|\\
= & \frac{n^2q_n}{4}\Abs{\sum_{(k_1,\dots,k_4)\in\mathcal{A}\cap\mathcal{A}_{2,2}}\E(\mathcal{F}_{k_1 k_2}\mathcal{F}_{k_3 k_4}) + \sum_{(k_1,\dots,k_4)\in\mathcal{A}\cap\mathcal{A}_{3}} \E(\mathcal{F}_{k_1 k_2}\mathcal{F}_{k_3 k_4})  }\\
= &O\bigg\{(10\omega_n^2M^2 + 24\omega_n M^3)q_n\tr(\bOme_{n,M}^2)\bigg\}\\
= &o(n^2\tr(\bOme_{n,M}^2)).
\end{align*}
This completes the proof.
\end{Prof}

\begin{Lem}\label{Lem for NG to G 3}
Under Assumptions 1--3 and $H_0$ holds, as $n \to \infty$,
\[\Delta S_3 = o_p(\tr^{1/2}(\bOme_{n,M}^2)).\]
\end{Lem}
\begin{Prof}
We partition $\Delta S_3$ into
\[\Delta S_{3,1} = \sum_{1\leq i,j\leq q_n}D_{ij}^{(G)} -  \sum_{1\leq i,j\leq q_n}D_{ij}^{(NG)}, \quad
 \Delta S_{3,2}= \sum_{j = 1}^{q_n} \sum_{i = j\omega_n - M+1}^{j\omega_n}\mathcal{F}_{ii}.\]
It's easy to prove that $\Delta S_{3,2} = o_p(\tr^{1/2}(\bOme_{n,M}^2))$. We will focus on $\Delta S_{3,1}$. Define
\begin{align*}
L_{ij} = \sum_{k = i\omega_n-M+1}^{i\omega_n}\sum_{l = (j-1)\omega_n+1}^{j\omega_n - M} \mathcal{F}_{kl},\\
R_{ij} = \sum_{k = (i-1)\omega_n+1}^{i\omega_n - M} \sum_{l = j\omega_n-M+1} ^ {j\omega_n}\mathcal{F}_{kl},\\
C_{ij} = \sum_{k = i\omega_n-M+1} ^ {i\omega_n} \sum_{l = j\omega_n-M+1} ^ {j\omega_n}\mathcal{F}_{kl}.
\end{align*}
Then, $D_{ij} = L_{ij} + R_{ij} + C_{ij}$. Since $\mathcal{F}_{kl} = \mathcal{F}_{lk}$, we can find out that  $R_{ij} = L_{ji}$. Then,
\begin{align*}
S_{3, 1}=&\sum_{1\leq i,j\leq q_n}2L_{ij}+\sum_{1\leq i,j\leq q_n}C_{ij}\\
=&2\bigg(\sum_{i = 1}^{q_n-1}\sum_{j=i+2}^{q_n} + \sum_{i = 2}^{q_n}\sum_{j = 1}^{i-1}\bigg)L_{ij} + \sum_{i = 1}^{q_n}L_{ii}
+ \sum_{i = 1}^{q_n-1}L_{i,i+1}+\bigg(2\sum_{i = 1}^{q_n}\sum_{j = i+1}^{q_n}+\sum_{i = 1}^{q_n}\bigg)C_{ij}\\
=& 2\mathcal{L}_1 + 2\mathcal{L}_2 + 2\mathcal{L}_3 + 2\mathcal{L}_4 + 2\mathcal{C}_1 + \mathcal{C}_2,
\end{align*}
where $\mathcal{L}_{1}=\sum_{i = 1}^{q_n-1}\sum_{j=i+2}^{q_n}L_{ij}^{(G)}$, $\mathcal{L}_{2}=\sum_{i = 2}^{q_n}\sum_{j = 1}^{i-1}L_{ij}^{(G)}$, $\mathcal{L}_{3}=\sum_{i = 1}^{q_n}L_{ii}^{(G)}$, $\mathcal{L}_{4}=\sum_{i = 1}^{q_n-1}L_{i,i+1}^{(G)}$, $\mathcal{C}_1=\sum_{i = 1}^{q_n}\sum_{j = i+1}^{q_n}C_{ij}^{(G)}$, $\mathcal{C}_2=\sum_{i = 1}^{q_n}C_{ij}^{(G)}$.
\begin{align*}
&\E[(\mathcal{L}_1^{(NG)})^2]\\
=&\sum_{i_1=1}^{q_n-1}\sum_{j_1=i_1+2}^{q_n}\sum_{i_2=1}^{q_n-1}\sum_{j_2=i_2+2}^{q_n}\sum_{k_1 = i_1\omega_n-M+1}^{i_1\omega_n}\sum_{k_2 = (j_1-1)\omega_n+1}^{j_1\omega_n-M}\sum_{k_3 = i_2\omega_n-M+1}^{i_2\omega_n}\sum_{k_4 = (j_2-1)\omega_n+1}^{j_2\omega_n-M}\E(\bgamma_{k_1}^\top \bgamma_{k_2}\bgamma_{k_3}^\top \bgamma_{k_4})\\
=&n^{-2}\sum_{i = 1}^{q_n-1} \sum_{j =i+2}^{q_n} \E\Brab{ \Brac{\Braa{\sum_{k = i\omega_n-M+1}^{i\omega_n}\bgamma_k}^\top \Braa{\sum_{l = (j-1)\omega_n+1}^{j\omega_n-M} \bgamma_l}}^2}\\
=&n^{-2}(q_n - 1)(q_n-2)M(\omega_n-M)\tr(\bOme_{M,M}\bOme_{\omega_n-M,M})\\
\leq&n^{-2}(q_n - 1)M[(q_n-2)(\omega_n-M)]\tr^{1/2}(\bOme_{M,M}^2)\tr^{1/2}(\bOme_{\omega_n-M,M}^2)\\
=&O(n^{-1}q_nM\tr(\bOme_{n,M}^2)) = o(\tr(\bOme_{n,M}^2)),
\end{align*}
the second equation holds due to the following fact. Firstly, $\E(\bgamma_{k_1}^\top \bgamma_{k_2}\bgamma_{k_3}^\top \bgamma_{k_4})\neq 0$ only if $k_{(2)} - k_{(1)}\leq M$ and $k_{(4)} - k_{(3)}\leq M$, where $k_{(1)}<\cdots<k_{(4)}$ be the sorted $k_1,\cdots,k_4$. Now, we consider the terms in the second equation.
Note that $j_1\geq i_1+2$ and $j_2\geq i_2+2$, then $k_2 - k_1 > M$ and $k_4 - k_3 > M$. Thus, $\E(\bgamma_{k_1}^\top \bgamma_{k_2}\bgamma_{k_3}^\top \bgamma_{k_4}) \neq 0$ only if $i_1= i_2$ or $i_1 = j_2$ or $i_1 = j_2-1$. When $i_1 = i_2$, $\E(\bgamma_{k_1}^\top \bgamma_{k_2}\bgamma_{k_3}^\top \bgamma_{k_4}) \neq 0$ only if $j_1 = j_2$. When $i_1 = j_2$ or $i_1 = j_2-1$, then $i_2\leq j_2 - 2 \leq i_1-1$, which implies $k_{(2)} - k_{(1)} = \min\{k_1- k_3, k_4 - k_3\} > M$. Therefore, $\E(\bgamma_{k_1}^\top \bgamma_{k_2}\bgamma_{k_3}^\top \bgamma_{k_4}) \neq 0$ only if $i_1 = i_2$ and $j_1 = j_2$.
Now, we have proved $\mathcal{L}_1^{(NG)} = o(\tr^{1/2}(\bOme_{n,M}^2))$. Similarly, we can prove that $\mathcal{L}_2^{(NG)} = o(\tr^{1/2}(\bOme_{n,M}^2))$, $\mathcal{L}_2^{(G)} = o(\tr^{1/2}(\bOme_{n,M}^2))$, $\mathcal{L}_2^{(G)} = o(\tr^{1/2}(\bOme_{n,M}^2))$, $\mathcal{C}_1^{(NG)} = o(\tr^{1/2}(\bOme_{n,M}^2))$ and $\mathcal{C}_1^{(G)} = o(\tr^{1/2}(\bOme_{n,M}^2))$.
\begin{align*}
\E[(\mathcal{L}_3^{(NG)})^2]&=\E \Brab{\Braa{\sum_{i = 1}^{q_n} \sum_{k = i\omega_n-M+1}^{i\omega_n}\sum_{l = (i-1)\omega_n+1}^{i\omega_n-M} \mathcal{F}_{kl}^{(NG)}}^2}\\
& = \sum_{i = 1}^{q_n}\E \Brab{\Braa{ \sum_{k = i\omega_n-M+1}^{i\omega_n}\sum_{l = (i-1)\omega_n+1}^{i\omega_n-M} \mathcal{F}_{kl}^{(NG)}}^2}\\
& = O(n^{-2}q_n (\omega_n-M)^2M^2\tr(\bOme_{n,M}^2)) = o(\tr(\bOme_{n,M}^2)).
\end{align*}
Similarly, we can prove that $\mathcal{L}_4^{(NG)} = o(\tr^{1/2}(\bOme_{n,M}^2))$, $\mathcal{L}_3^{(G)} = o(\tr^{1/2}(\bOme_{n,M}^2))$, $\mathcal{L}_4^{(G)} = o(\tr^{1/2}(\bOme_{n,M}^2))$, $\mathcal{C}_2^{(NG)} = o(\tr^{1/2}(\bOme_{n,M}^2))$ and $\mathcal{C}_2^{(G)} = o(\tr^{1/2}(\bOme_{n,M}^2))$. This completes the proof.
\end{Prof}

\begin{Lem}\label{Lem for NG to G 4}
Under Assumptions 1--3 and $H_0$ holds, as $n \to \infty$,
\[\Delta S_4 = o_p(\tr^{1/2}(\bOme_{n,M}^2)).\]
\end{Lem}
\begin{Prof} We partition
\[\Delta S_4 = F^{(G)} - F^{(NG)} + \sum_{k = q_n\omega_n+1}^n\mathcal{F}^{(NG)}_{kk}- \sum_{k = q_n\omega_n+1}^n\mathcal{F}^{(G)}_{kk}
\defeq \Delta S_{4,1} + \Delta S_{4,2}.\]
Since $\Delta S_{4,2} = o_p(\tr^{1/2}(\bOme_{n,M}^2))$ is trivial, it suffices to prove $\Delta S_{4,1} = o_p(\tr^{1/2}(\bOme_{n,M}^2)).$
Define
\begin{align*}
\mathcal{B} &= \{(k_1,k_2,k_3,k_4)\in \{1,\cdots,n\}\times\{q_n\omega_n+1,\cdots,n\}\times \{1,\cdots,n\}\times\{q_n\omega_n+1,\cdots,n\}\} , \\
\mathcal{B}_1 &= \{(k_1,k_2,k_3,k_4)\in \{1,\cdots,n\}^4:\Abs{k_1 - k_3}\geq 3M;\  \min(k_1,k_3) > q_n\omega_n\}, \\
\mathcal{B}_2 &= \{(k_1,k_2,k_3,k_4)\in \{1,\cdots,n\}^4:\Abs{k_1 - k_3}\geq 3M;\  \min(k_1,k_3) > q_n\omega_n;\\
&\quad\quad\min(\Abs{k_1-k_2},\Abs{k_1-k_4}) \leq M; \min(\Abs{k_3-k_2},\Abs{k_3-k_4}) \leq M\}.
\end{align*}
For any $(k_1,k_2,k_3,k_4) \in \mathcal{B}_1\setminus\mathcal{B}_2$, we have $\E(\mathcal{F}_{k_1k_2}\mathcal{F}_{k_3k_4}) = 0$. Note that $\mathcal{B}_2 \subset \{q_n\omega_n+1,\cdots,n\}^4$, then $|\mathcal{A}_2|\leq r_n^4$. Since $|\mathcal{B}\setminus\mathcal{B}_1|\leq 6Mn\omega_n^2$, by Corollary \ref{Coro for fourth moment of gamma}, we have
\begin{align*}
\E[(F^{(NG)})^2]&= \E\Brab{\Braa{\sum_{k = 1}^n\sum_{l = q_n\omega_n + 1}^n\mathcal{F}_{kl} + \sum_{k = q_n\omega_n + 1}^n\sum_{l = 1}^n\mathcal{F}_{kl}}^2}  \\
&\leq 4\sum_{k = 1}^n\sum_{l = q_n\omega_n + 1}^n\sum_{r =1}^n\sum_{s = q_n\omega_n + 1}^n\Abs{\E(\mathcal{F}_{kl}\mathcal{F}_{rs})}\\
& = 4\sum_{(k,l,r,s) \in \mathcal{A}\setminus \mathcal{A}_1} \Abs{\E(\mathcal{F}_{kl}\mathcal{F}_{rs})}+4\sum_{(k,l,r,s) \in \mathcal{A}_2} \Abs{\E(\mathcal{F}_{kl}\mathcal{F}_{rs})}\\
& = O\bigg\{ \frac{6Mn\omega_n^2 + r_n^4}{n^2}\tr\Braa{\bOme_{n,M}^2}\bigg\} = o(\tr(\bOme_{n,M}^2)).
\end{align*}
Similarly, we have $\E[(F^{(G)})^2]=o(\tr(\bOme_{n,M}^2))$. This completes the proof.
\end{Prof}

\subsubsection{Part 3}
\begin{Lem}\label{Lem in part3}
Under Assumptions 1--3 and $H_0$ holds, as $n \to \infty$,
\[T_n - T_n^{(NG)} = o_p(\tr^{1/2}(\bOme_{n,M}^2)).\]
\end{Lem}
\begin{Prof}
We reformulate
\begin{align*}
T_n - T_n^{(NG)} = &\frac1n\sum_{t = 2}^n\sum_{s = 1}^{t-1}(\bepsilon_t^\top\bepsilon_s - \bgamma_t^\top\bgamma_s)\\
= & \frac1n\sum_{t = 2}^n\sum_{s = 1}^{t-1}(\bepsilon_t - \bgamma_t)^\top(\bepsilon_s - \bgamma_s) + \frac1n\sum_{1\leq t, s\leq n,t\neq s}\bgamma_t^\top(\bepsilon_s - \bgamma_s).
\end{align*}
By the calculations, we have
\begin{align*}
&\Var\bigg\{\sum_{1\leq t, s\leq n, t\neq s}\bgamma_t^\top(\bepsilon_s - \bgamma_s)\bigg\}\\
= & \sum_{\substack{1\leq t_1,t_2,s_1,s_2\leq n\\t_1\neq s_1,t_2\neq s_2}}\sum_{\substack{t_1-M\leq u_1 \leq t_1\\t_2-M\leq u_2\leq t_2}}\sum_{\substack{v_1 \leq s_1-M-1\\v_2\leq s_2-M-1}} b_{t_1 - u_1}b_{s_1 - v_1}b_{t_2 - u_2}b_{s_2 - v_2}\\
&\quad\quad\quad\quad\quad\quad\quad\quad\quad\quad
\times\E[(\bZ_{u_1}^\top\bSig \bZ_{v_1} - \delta_{u_1,v_1}\tr(\bSig))(\bZ_{u_2}^\top\bSig \bZ_{v_2} - \delta_{u_2,v_2}\tr(\bSig))]\\
= & \sum_{\substack{1\leq t_1,t_2,s_1,s_2\leq n\\t_1\neq s_1,t_2\neq s_2}}\Bigg\{
\sum_{\substack{t_1-M\leq r\leq t_1\land(s_1-M-1)\\t_2-M\leq s\leq t_2 \land(s_2-M-1), r\neq s}} b_{t_1 - r}b_{s_1 - r}b_{t_2 - s}b_{s_2 - s}\E[(\bZ_{r}^\top\bSig \bZ_{r} - \tr(\bSig))(\bZ_{s}^\top\bSig \bZ_{s} - \tr(\bSig))]\\
& + \sum_{\substack{(t_1-M)\land(t_2-M)\leq r\leq t_1\land t_2\\s\leq (s_1-M-1)\land(s_2-M-1), r\neq s}} b_{t_1 - r}b_{s_1 - s}b_{t_2 - r}b_{s_2 - s}\E[(\bZ_{r}^\top\bSig \bZ_{s})(\bZ_{r}^\top\bSig \bZ_{s})]\\
& + \sum_{\substack{t_1-M\leq r\leq t_1\land(s_2-M-1)\\t_2-M\leq s\leq t_2 \land(s_1-M-1), r\neq s}} b_{t_1 - r}b_{s_1 - s}b_{t_2 - s}b_{s_2 - r}\E[(\bZ_{r}^\top\bSig \bZ_{s})(\bZ_{s}^\top\bSig \bZ_{r})]\\
& + \sum_{\substack{(t_1-M)\land(t_2-M)\leq r \leq t_1\land t_2\\
\land (s_1-M-1)\land(s_2-M-1)}}b_{t_1 - r}b_{s_1 - r}b_{t_2 - r}b_{s_2 - r}\E[(\bZ_{r}^\top\bSig \bZ_{r} - \tr(\bSig))(\bZ_{r}^\top\bSig \bZ_{r} -\tr(\bSig))]\Bigg\}\\
\leq & \sum_{\substack{1\leq t_1,t_2,s_1,s_2\leq n\\t_1\neq s_1,t_2\neq s_2}}\Bigg\{\sum_{\substack{(t_1-M)\land(t_2-M)\leq r\leq t_1\land t_2\\s\leq (s_1-M-1)\land(s_2-M-1), r\neq s}}
  \Abs{b_{t_1 - r}b_{t_2 - r}}\Abs{b_{s_1 - s}b_{s_2 - s}}\tr(\bSig^2)\\
  & + \sum_{\substack{t_1-M\leq r\leq t_1\land(s_2-M-1)\\t_2-M\leq s\leq t_2 \land(s_1-M-1), r\neq s}}\Abs{b_{t_1 - r}b_{s_2 - r}}\Abs{b_{s_1 - s}b_{t_2 - s}}\tr(\bSig^2) \\
& + \sum_{(t_1-M)\land(t_2-M)\leq r \leq t_1\land t_2\land (s_1-M-1)\land(s_2-M-1)}\Abs{b_{t_1 - r}b_{t_2 - r}b_{s_1 - r}b_{s_2 - r}}\tau_1 \tr^2(\bSig)\Bigg\}\\
\leq &\Bigg\{\Braa{n\sum_{t_1,t_2\geq 0}\Abs{b_{t_1}b_{t_2}}}\Braa{n\sum_{s_1,s_2\geq M+1}\Abs{b_{s_1}b_{s_2}}} + \Bigg(n\sum_{t_1\geq 0, s_2\geq M+1}\Abs{b_{t_1}b_{s_2}}\Bigg)\Bigg(n\sum_{t_2\geq 0, s_1\geq M+1}\Abs{b_{s_1}b_{s_2}}\Bigg)
\end{align*}
\begin{align*}
& + \tau_1pn\sum_{t_1,t_2\geq 0}\sum_{s_1,s_2\geq M+1}\Abs{b_{t_1}b_{t_2}b_{s_1}b_{s_2}}\Bigg\}\tr(\bSig^2) = o(n^2\tr(\bOme_{n,M}^2)).
\end{align*}
The last inequality holds due to Lemma \ref{Lem for the order of bn}.  Similarly, we can prove that
\[\Var\bigg\{\frac1n\sum_{i = 2}^n\sum_{j = 1}^{i-1}(\bepsilon_i - \bgamma_i)^\top(\bepsilon_j - \bgamma_j)\bigg\} = o(n^2\tr(\bOme_{n,M}^2)) \]
This completes the proof.
\end{Prof}

\begin{Lem}\label{Lem in part3}
Under Assumptions 1--3 and $H_0$ holds, as $n \to \infty$,
\begin{equation}
\frac{1}{2}\tr(\bOme_{n,M}^2)  = \sigma_n^2(1+ o(1)).
\end{equation}
\end{Lem}
\begin{Prof}
By some calculations, we have
\begin{align*}
\bOme - \bOme_{n,M}  = & \bigg[\bigg\{a_0+2\sum\limits_{h=1}^{\infty}a_h\bigg\}-\bigg\{a_{0,M}+2\sum\limits_{h=1}^{M}\bigg(1-\frac{h}{n}\bigg)a_{h,M}\bigg\}\bigg]\bSig.
\end{align*}
By
\begin{align*}
&\Abs{\bigg\{a_0+2\sum\limits_{h=1}^{\infty}a_h\bigg\}-\bigg\{a_{0,M}+2\sum\limits_{h=1}^{M}\bigg(1-\frac{h}{n}\bigg)a_{h,M}\bigg\}}\\
\leq & 2\Abs{\sum_{h = M+1}^{\infty} \bigg(1-\frac{h}{n}\bigg) \sum_{i = 0}^\infty b_ib_{i+h} + \sum_{h = 0}^M \bigg(1-\frac{h}{n}\bigg) \sum_{i = M-h+1}^\infty b_ib_{i+h} +
\sum_{h = 0}^M\frac{h}{n} \sum_{i = 0}^\infty b_ib_{i+h}}\\
\leq & 4\bigg(\sum_{i = 0}^\infty \Abs{b_i}\bigg)\bigg(\sum_{i = M+1}^\infty \Abs{b_i}\bigg) + n^{-1}M\sum_{h = 0}^M\sum_{i = 0}^\infty b_ib_{i+h}\\
= & o(M^{-4}),
\end{align*}
We have $\tr\{(\bOme - \bOme_{n,M})^2\} = o(M^{-8})\tr(\bSig^2)$ and $\tr\{\bOme_{n,M}(\bOme - \bOme_{n,M})\} = o(M^{-4})\tr(\bSig^2)$.
Furthermore, by $\tr(\bOme^2) - \tr(\bOme_{n,M}^2) = \tr\{(\bOme - \bOme_{n,M})^2\} + 2\tr\{\bOme_{n,M}(\bOme - \bOme_{n,M})\}$ and $\tr(\bOme^2) = s^4\tr(\bSig^2)$, we have
\begin{equation*}
\frac{1}{2}\tr(\bOme_{n,M}^2) = \sigma_n^2(1+ o(1)).
\end{equation*}
\end{Prof}

\subsection{Proof of Theorem \ref{th22}}
Note that
\begin{align*}
\frac{T_n - \hat{\mu}_n}{\sqrt{\hat{\sigma}^2_n}}=\bigg(\frac{T_n - \mu_n}{\sqrt{{\sigma}^2_n}}+\frac{\mu_n - \hat{\mu}_n}{\sqrt{{\sigma}^2_n}}\bigg)\cdot\sqrt{\frac{{\sigma}^2_n}{\hat{\sigma}^2_n}}
\end{align*}
It is sufficient to prove $(\hat{\mu}_n - \mu_{n})/\sqrt{\sigma_n^2}\limP 0$ and $\hat{\sigma}^2_n / \sigma^2_n\limP 1$.

\subsubsection{Part 1}
\begin{Prof}
In this part, we prove $(\hat{\mu}_n - \mu_{n})/\sqrt{\sigma_n^2}\limP 0$.
By the calculations, we have
\begin{align*}
\hat{\mu}_n = \frac{1}{n^2}\sum_{h = 1}^{M}\sum_{t = 1}^{n-h}(n-h)(\bepsilon_t^\top\bepsilon_{t+h}-\bepsilon_t^\top\bar{\bepsilon}_n-\bepsilon_{t+h}^\top\bar{\bepsilon}_n+\bar{\bepsilon}_n^\top\bar{\bepsilon}_n).
\end{align*}
Define $$\mu_n^{*}=\sum_{h = 1}^{M}\sum_{t = 1}^{n-h}\bepsilon_t^\top\bepsilon_{t+h}/n,$$
it is equivalent to prove
$$\mu_n^{*}-\mu_{n}=o_p(\tr^{1/2}(\bOme_n^2)) \quad \mbox{and} \quad  \hat{\mu}_n -\mu_n^{*}=o_p(\tr^{1/2}(\bOme_n^2)).$$
$\E(\mu_n^{*})=\mu_{n}$ and
\begin{align*}
\Var(\mu_n^{*}) =&\frac1{n^2}\sum_{h_1 = 1}^M\sum_{h_2 = 1}^M\sum\limits_{i_1 = 1}^{n-h_1}\sum\limits_{i_2 = 1}^{n-h_2}
\E[\{\bepsilon_{i_1}^\top\bepsilon_{i_1+h_1} - \E(\bepsilon_{i_1}^\top\bepsilon_{i_1+h_1})\}\{\bepsilon_{i_2}^\top\bepsilon_{i_2+h_2} - \E(\bepsilon_{i_2}^\top\bepsilon_{i_2+h_2})\}]\\
= &\frac1{n^2}\sum_{h_1 = 1}^M\sum_{h_2 = 1}^M\sum\limits_{t_1 = 1}^{n-h_1}\sum\limits_{t_2 = 1}^{n-h_2}\E\Bigg[\bigg\{\sum_{u_1\leq t_1}\sum_{u_2\leq i_1+ h_1}b_{u_1}b_{u_2}\bigg(\bZ_{i_1 - u_1}^\top \bSig \bZ_{i_1 + h_1 - u_2} - \delta_{i_1-u_1,i_1+h_1 - u_2}\tr(\bSig)\bigg)\bigg\}\\
& \bigg\{\sum_{u_3\leq i_2}\sum_{u_4\leq i_2+ h_2}b_{u_3}b_{u_4}\bigg(\bZ_{i_2 - u_3}^\top\bSig\bZ_{i_2 + h_2 - u_4} - \delta_{i_2-u_3,i_2+h_2 - u_4}\tr(\bSig)\bigg)\bigg\}\Bigg]\\
= &\frac1{n^2}\sum_{h_1 = 1}^M\sum_{h_2 = 1}^M\sum\limits_{i_1 = 1}^{n-h_1}\sum\limits_{i_2 = 1}^{n-h_2}\sum_{j_1=1}^{i_1}\sum_{j_2=1}^{i_1+ h_1}\sum_{j_3=1}^{i_2}\sum_{j_4=1}^{i_2+ h_2}b_{i_1-j_1}b_{i_1+h_1-j_2}b_{i_2-j_3}b_{i_2+h_2-j_4}\\
&\E\Bigg[\bigg\{\bZ_{j_1}^\top \bSig \bZ_{j_2} - \delta_{j_1,j_2}\tr(\bSig)\bigg\}\bigg\{\bZ_{j_3}^\top \bSig \bZ_{j_4} - \delta_{j_3,j_4}\tr(\bSig)\bigg\}\Bigg].
\end{align*}
Consider the value of $\rho(j_1,j_2,j_3,j_4)\defeq \E[(\bZ_{j_1}^\top \bSig \bZ_{j_2} - \delta_{j_1,j_2}\tr(\bSig))(\bZ_{j_3}^\top \bSig \bZ_{j_4} - \delta_{j_3,j_4}\tr(\bSig))]$.
\[\rho(j_1,j_2,j_3,j_4)\begin{cases}
    = \tr(\bSig^2), & j_1 = j_3= r, j_2= j_4 =s, r\neq s\\
    = \tr(\bSig^2), & j_1 = j_4 = r, j_2= j_3 = s, r\neq s\\
    \leq \tau_1\tr^2(\bSig), & j_1 = j_2 = j_3 = j_4 = r\\
    = 0, & \text{otherwise}.
\end{cases}\]
So, we need to consider the coefficient
\[C_{j_1,j_2,j_3,j_4}\defeq\sum_{j_1=1}^{i_1}\sum_{j_2=1}^{i_1+ h_1}\sum_{j_3=1}^{i_2}\sum_{j_4=1}^{i_2+ h_2}b_{i_1-j_1}b_{i_1+h_1-j_2}b_{i_2-j_3}b_{i_2+h_2-j_4}.\]
Then,
\[\sum_{i_1 = 1}^{n-h_1}\sum_{i_2 = 1}^{n-h_2}\Abs{C_{r,s,r,s}} \leq \sum_{i_1 = 1}^{n-h_1}\sum_{i_2 = 1}^{n-h_2}\sum_{r=1}^{i_1\wedge i_2}\sum_{s=1}^{(i_1+h_1)\wedge (i_2+h_2)}|b_{i_1-r}b_{i_1+h_1-s}b_{i_2-r}b_{i_2+h_2-s}|\]
Consider the number of $\Abs{b_{k_1}b_{k_2}b_{k_3}b_{k_4}}$ appear in the right hand side. This can be equivalently written as
\[i_1 - r = k_1\quad i_1 + h_1 - s = k_2 \quad i_2 - r = k_3\quad i_2 + h_2 - s = k_4.\]
Fix $i_1$ and $k_1,k_2,k_3,k_4$, $(r,s,i_2)$ can be uniquely represented by $(i_1,k_1,k_2,k_3,k_4)$, or not exist. Therefore, $\Abs{b_{k_1}b_{k_2}b_{k_3}b_{k_4}}$ appears at most $n-h_1$ times. We can conclude that
\[\sum_{r\leq n-h_1}\sum_{s\leq n-h_2}\Abs{C_{r,s,r,s}}\leq n\bigg(\sum_{i = 0}^\infty \Abs{b_i}\bigg)^4\]
Similarly, we can prove that
\[\sum_{r\leq n-h_1}\sum_{s\leq n-h_2}\Abs{C_{r,s,s,r}}\leq n\bigg(\sum_{i = 0}^\infty \Abs{b_i}\bigg)^4 \quad \sum_{r\leq n- h_1\land h_2}\Abs{C_{r,r,r,r}}\leq \bigg(\sum_{i = 0}^\infty \Abs{b_i}\bigg)^4\]
By $\tr^2(\Sigma)\leq p\tr(\Sigma^2)$, we have
\begin{eqnarray*}
\var(\mu_n^{*})  =O(M^2(n^{-1} + n^{-2}p)\tr(\bSig^2)).
\end{eqnarray*}
Thus, $\mu_n^{*}-\mu'_{n}=o_p(\tr^{1/2}(\bOme_{n}^2))$. Now, we only need to prove $\hat{\mu}_n -\mu_n^{*}=o_p(\tr^{1/2}(\bOme_n^2))$. We mainly focus on proving
\[\frac{1}{n^2}\sum_{h = 1}^{M}\sum_{t = 1}^{n-h}(n-h)\bigg\{\bepsilon_t^\top\bar{\bepsilon}_n+\bepsilon_{t+h}^\top\bar{\bepsilon}_n-\bar{\bepsilon}_n^\top\bar{\bepsilon}_n\bigg\}=o_p(\tr^{1/2}(\bOme_n^2)).\]
Since $\sum_{t_1 = 1}^{n-h_1}\sum_{t_2 = 1}^{n-h_2}\sum_{j=1}^{t_1\wedge t_2}|b_{t_1-j}b_{t_2-j}|\leq n(\sum_{i = 0}^\infty|b_i|)^2$, under Assumptions (A1)--(A3) and $M = \lceil \min(n,p)^{1/8}\rceil $,
\begin{align*}
&\E\bigg\{\frac{1}{n^4}\sum_{h_1 = 1}^{M}\sum_{t_1 = 1}^{n-h_1}\sum_{h_2 = 1}^{M}\sum_{t_2 = 1}^{n-h_2}(n-h_1)(n-h_2)\bepsilon_{t_1}^\top\bepsilon_{t_2}\bigg\}\\
=&\E\bigg\{\frac{1}{n^4}\sum_{h_1 = 1}^{M}\sum_{t_1 = 1}^{n-h_1}\sum_{h_2 = 1}^{M}\sum_{t_2 = 1}^{n-h_2}\sum\limits_{j_1=1}^{t_1}\sum\limits_{j_2=1}^{t_2}(n-h_1)(n-h_2)b_{t_1-j_1}b_{t_2-j_2}Z_{j_1}^\top\Sigma Z_{j_2}\bigg\}\\
=&\frac{1}{n^4}\sum\limits_{h_1 = 1}^{M}\sum\limits_{h_2 = 1}^{M}(n-h_1)(n-h_2)\sum_{t_1 = 1}^{n-h_1}\sum_{t_2 = 1}^{n-h_2}\sum\limits_{j=1}^{t_1\wedge t_2}b_{t_1-j}b_{t_2-j}\tr(\bSig),\\
=&O(n^{-1}p^{1/2}M^2\tr^{1/2}(\bOme_n^2))=o(\tr^{1/2}(\bOme_n^2)),
\end{align*}
and
\begin{align*}
\E(\bar{\bepsilon}_n^\top\bar{\bepsilon}_n)=\frac{1}{n}\tr(\bOme_n)\leq \frac{\sqrt{p}}{n}\tr^{1/2}(\bOme_n^2)=o(\tr^{1/2}(\bOme_n^2)).
\end{align*}
These implies $\hat{\mu}_n - \mu'_{n} = o_p(\tr^{1/2}(\bOme_n^2))$.

\subsubsection{Part 2}
In this part, we prove $\hat{\sigma}^2_n / \sigma^2_n\limP 1$. Define $\hat{\sigma}^2_{n}=\hat{\sigma}^2_{n,1}+\hat{\sigma}^2_{n,2}$, where
\begin{align*}
\hat{\sigma}^2_{n,1} = &\frac12\bigg(S_{0,0,1} + 2\sum_{r = 1}^M S_{r,0,1} + 2\sum_{r = 1}^M S_{0,r,1} + 4\sum_{r = 1}^M\sum_{s = 1}^M S_{r,s,1}\bigg),\\
\hat{\sigma}^2_{n,2} = &\frac12\bigg(S_{0,0,2} + 2\sum_{r = 1}^M S_{r,0,2} + 2\sum_{r = 1}^M S_{0,r,2} + 4\sum_{r = 1}^M\sum_{s = 1}^M S_{r,s,2}\bigg),\\
S_{r,s,1} = &\frac{\displaystyle\sum_{i = 1}^{[n/2]-s}\displaystyle\sum_{j = i + [n/2]}^{n-s}\bepsilon_{i}^\top\bepsilon_j\bepsilon_{i+r}^\top\bepsilon_{j+s}}{(n-s/2-\frac32 [n/2] +1/2)([n/2]-s)}, \quad S_{r,s,2} = S_{r,s}-S_{r,s,1}.
\end{align*}
We will prove $\hat{\sigma}^2_{n,1}/\sigma_n^2\limP 1$ and $\hat{\sigma}^2_{n,2}/\sigma_n^2\limP 0$.

{\noindent\textbf{Step 1.}} In this step, we prove $\hat{\sigma}^2_{n,1}/\sigma_n^2\limP 1$. It suffice to prove $(S_{r,s,1}-a_{r}a_{s}\tr(\bSig^2))/\tr(\bSig^2)=o_p(1)$, which implies that
\begin{align*}
\frac{S_{0,0,1} + 2\sum_{r = 1}^M S_{r,0,1} + 2\sum_{r = 1}^M S_{0,r,1} + 4\sum_{r = 1}^M\sum_{s = 1}^M S_{r,s,1}}{(a_0^2+4\sum_{r = 1}^Ma_0a_r+4\sum_{r = 1}^M\sum_{s = 1}^Ma_ra_s)\tr(\bSig^2)}=1+o(1).
\end{align*}
Note that
\begin{align*}
&(n-s/2-\frac32 [n/2] +1/2)([n/2]-s)\E(S_{r,s,1}-a_ra_s\tr(\bSig^2))\\
=&\displaystyle\sum_{\substack{1\leq i \leq [n/2]-s\\i + [n/2]\leq j \leq n-s}}\bigg\{\sum\limits_{\substack{\ell_1\leq i,\ell_2\leq j \\ \ell_3\leq i+r,\ell_4\leq j+s}}b_{i-\ell_1}b_{j-\ell_2}b_{i+r-\ell_3}b_{j+s-\ell_4}\E(\bZ_{\ell_1}^\top\bSig \bZ_{\ell_2}\bZ_{\ell_3}^\top\bSig \bZ_{\ell_4})-a_ra_s\tr(\bSig^2)\bigg\}\\
\leq&\displaystyle\sum_{\substack{1\leq i \leq [n/2]-s\\i + [n/2]\leq j \leq n-s}}\bigg\{\bigg(\sum\limits_{\substack{\ell_1\leq i,\ell_2\leq j}}b_{i-\ell_1}b_{j-\ell_2}b_{i+r-\ell_1}b_{j+s-\ell_2}
+\sum\limits_{\substack{\ell_1\leq i\wedge (j+s)\\\ell_2\leq j\wedge (i+r)}}b_{i-\ell_1}b_{j-\ell_2}b_{i+r-\ell_2}b_{j+s-\ell_1}\bigg)\tr(\bSig^2)\\
&+\sum\limits_{\substack{\ell_1\leq i\wedge j\\\ell_2\leq (i+r)\wedge(j+s)}}b_{i-\ell_1}b_{j-\ell_1}b_{i+r-\ell_2}b_{j+s-\ell_2}\tr^2(\bSig)+Cp\sum\limits_{\ell\leq i\wedge j}b_{i-\ell}b_{j-\ell}b_{i+r-\ell}b_{j+s-\ell-a_ra_s}\tr(\bSig^2)\\
&-a_ra_s\tr(\bSig^2)\bigg\}\\
=&O\bigg[\displaystyle\sum_{\substack{1\leq i \leq [n/2]-s\\i + [n/2]\leq j \leq n-s}}\bigg\{\bigg(a_{j+s-i}a_{i+r-j}+pa_{j-i}a_{j+s-i-r}+p\sum\limits_{\ell\leq i\wedge j}b_{i-\ell}b_{j-\ell}b_{i+r-\ell}b_{j+s-\ell-a_ra_s}\bigg)\tr(\bSig^2)\bigg\}\bigg]
\end{align*}
\begin{align*}
=&o(n^{-1}p\tr(\bSig^2)).
\end{align*}
Hence, $\E\{(S_{r,s,1}-a_ra_s\tr(\bSig^2))/\tr(\bSig^2)\}=o(n^{-3}p)=o(1)$. Similarly, we can verify that $\var\{(S_{r,s,1}-a_ra_s\tr(\bSig^2))/\tr(\bSig^2)\}=o(pn^{-2}+n^{-1})=o(1)$. Thus,
$(S_{r,s,1}-a_ra_s\tr(\bSig^2))/\tr(\bSig^2)=o_p(1)$. It follows that,
\begin{align*}
\frac{S_{0,0,1} + 2\sum_{r = 1}^M S_{r,0,1} + 2\sum_{r = 1}^M S_{0,r,1} + 4\sum_{r = 1}^M\sum_{s = 1}^M S_{r,s,1}}{(a_0^2+4\sum_{r = 1}^M a_oa_r + 4\sum_{r = 1}^M\sum_{s = 1}^Ma_ra_s)\tr(\bSig^2)}=1+o_p(1).
\end{align*}
Now, we get $\hat{\sigma}^2_{n,1}/\sigma_n^2\limP 1$.

{\noindent\textbf{Step 2.}}  In this step, we will prove $\hat{\sigma}^2_{n,2}/\sigma_n^2\limP 0$. It suffices to prove that $\sum_{r = 1}^M\sum_{s = 1}^MS_{r,s,2}/\tr(\bSig^2)=o_p(1)$.
\begin{align*}
&(\bX_{i}-\bar{\bX}_n)^\top(\bX_j-\bar{\bX}_n)(\bX_{i+r}-\bar{\bX}_n)^\top(\bX_{j+s}-\bar{\bX}_n)\\
=&(\bepsilon_{i}-\bar{\bepsilon}_n)^\top(\bepsilon_j-\bar{\bepsilon}_n)(\bepsilon_{i+r}-\bar{\bepsilon}_n)^\top(\bepsilon_{j+s}-\bar{\bepsilon}_n)\\
=&(\bepsilon_{i}^\top\bepsilon_j-\bepsilon_{i}^\top\bar{\bepsilon}_n-\bepsilon_{j}^\top\bar{\bepsilon}_n
+\bar{\bepsilon}_n^\top\bar{\bepsilon}_n)(\bepsilon_{i+r}^\top\bepsilon_{j+s}-\bepsilon_{i+r}^\top\bar{\bepsilon}_n-\bepsilon_{j+s}^\top\bar{\bepsilon}_n+\bar{\bepsilon}_n^\top\bar{\bepsilon}_n)
\end{align*}
We first verify that
$$\displaystyle\sum_{r = 1}^M\displaystyle\sum_{s = 1}^M\frac{\displaystyle\sum\nolimits_{\substack{1\leq i \leq [n/2]-s\\i + [n/2]\leq j \leq n-s}}\bepsilon_{i}^\top\bepsilon_j\bepsilon_{i+r}^\top\bar{\bepsilon}_n}{(n-s/2-\frac32 [n/2] +1/2)([n/2]-s)}=o_p(\tr(\bSig^2)).$$
Note that
\begin{align*}
&\displaystyle\sum_{\substack{1\leq i_1 \leq [n/2]-s\\i_1 + [n/2]\leq j_1 \leq n-s}}\displaystyle\sum_{\substack{1\leq i_2 \leq [n/2]-s\\i_2 + [n/2]\leq j_2 \leq n-s}}\E(\bepsilon_{i_1}^\top\bepsilon_{j_1}\bepsilon_{i_1+r}^\top\bepsilon_{i_2}^\top\bepsilon_{j_2}\bepsilon_{i_2+r}^\top)\\
=&\displaystyle\sum_{\substack{1\leq i_1 \leq [n/2]-s\\i_1 + [n/2]\leq j_1 \leq n-s}}\displaystyle\sum_{\substack{1\leq i_2 \leq [n/2]-s\\i_2 + [n/2]\leq j_2 \leq n-s}}\bigg\{\sum\limits_{\substack{\ell_1\leq i_1,\ell_2\leq j_1 \\ \ell_3\leq i_1+r}}\sum\limits_{\substack{\ell_4\leq i_2,\ell_5\leq j_2 \\ \ell_6\leq i_2+r}}b_{i_1-\ell_1}b_{j_1-\ell_2}b_{i_1+r-\ell_3}b_{i_2-\ell_4}b_{j_2-\ell_5}b_{i_2+r-\ell_6}\\
&\E(\bZ_{\ell_1}^\top\bSig \bZ_{\ell_2}\bZ_{\ell_3}^\top\bSig \bZ_{\ell_4}\bSig \bZ_{\ell_5}^\top\bSig \bZ_{\ell_6})\bigg\}.
\end{align*}
Define $\rho(\ell_1,\ell_2,\ell_3,\ell_4,\ell_5,\ell_6)=\E(\bZ_{\ell_1}^\top\bSig \bZ_{\ell_2}\bZ_{\ell_3}^\top\bSig \bZ_{\ell_4}\bSig \bZ_{\ell_5}^\top\bSig \bZ_{\ell_6})$,
\[\rho(\ell_1,\ell_2,\ell_3,\ell_4,\ell_5,\ell_6)\begin{cases}
    = \tr^3(\bSig), & \ell_1 = \ell_2= a, \ell_3=\ell_4=b, \ell_5=\ell_6=c, a\neq b \neq c\\
    = \tr(\bSig)\tr(\bSig^2), & \ell_1 = \ell_3= a, \ell_2=\ell_4=b, \ell_5=\ell_6=s, a\neq b \neq c\\
    \leq \tau_1\tr^3(\bSig), & \ell_1 = \ell_2= a, \ell_3=\ell_4=\ell_5=\ell_6=b, a\neq b\\
     \leq \tau_2\tr^3(\bSig), &\ell_1 = \ell_2= \ell_3=\ell_4=\ell_5=\ell_6=a\\
    = 0, & \text{otherwise}.
\end{cases}\]
Now, we consider the coefficient
\[C_{\ell_1,\ell_2,\ell_3,\ell_4,\ell_5,\ell_6}\defeq\sum\limits_{\substack{\ell_1\leq i_1,\ell_2\leq j_1 \\ \ell_3\leq i_1+r}}\sum\limits_{\substack{\ell_4\leq i_2,\ell_5\leq j_2 \\ \ell_6\leq i_2+r}}b_{i_1-\ell_1}b_{j_1-\ell_2}b_{i_1+r-\ell_3}b_{i_2-\ell_4}b_{j_2-\ell_5}b_{i_2+r-\ell_6}.\]
Similar to the discussion in the proof of proposition 2, we have
\begin{eqnarray*}
&&\displaystyle\sum_{\substack{1\leq i_1 \leq [n/2]-s\\i_1 + [n/2]\leq j_1 \leq n-s}}\displaystyle\sum_{\substack{1\leq i_2 \leq [n/2]-s\\i_2 + [n/2]\leq j_2 \leq n-s}}C_{a,a,b,b,c,c}\\
&\leq&\displaystyle\sum_{\substack{1\leq i_1 \leq [n/2]-s\\i_1 + [n/2]\leq j_1 \leq n-s}}\displaystyle\sum_{\substack{1\leq i_2 \leq [n/2]-s\\i_2 + [n/2]\leq j_2 \leq n-s}}\sum\limits_{\substack{a\leq i_1\wedge j_1,b\leq (i_1+r)\wedge  i_2 \\ c\leq j_2 \wedge (i_2+r)}}|b_{i_1-a}b_{j_1-a}b_{i_1+r-b}b_{i_2-b}b_{j_2-c}b_{i_2+r-c}|\\
&\leq& n\bigg(\sum_{i = 0}^\infty \Abs{b_i}\bigg)^6\\
&&\displaystyle\sum_{\substack{1\leq i_1 \leq [n/2]-s\\i_1 + [n/2]\leq j_1 \leq n-s}}\displaystyle\sum_{\substack{1\leq i_2 \leq [n/2]-s\\i_2 + [n/2]\leq j_2 \leq n-s}}C_{a,b,a,b,c,c},\\
&\leq&\displaystyle\sum_{\substack{1\leq i_1 \leq [n/2]-s\\i_1 + [n/2]\leq j_1 \leq n-s}}\displaystyle\sum_{\substack{1\leq i_2 \leq [n/2]-s\\i_2 + [n/2]\leq j_2 \leq n-s}}\sum\limits_{\substack{a\leq i_1,b\leq j_1\wedge  i_2 \\ c\leq j_2 \wedge (i_2+r)}}|b_{i_1-a}b_{j_1-b}b_{i_1+r-a}b_{i_2-b}b_{j_2-c}b_{i_2+r-c}|\\
&\leq& n^2\bigg(\sum_{i = 0}^\infty \Abs{b_i}\bigg)^6
\end{eqnarray*}
Furthermore, we have
\begin{eqnarray*}
\displaystyle\sum_{\substack{1\leq i_1 \leq [n/2]-s\\i_1 + [n/2]\leq j_1 \leq n-s}}\displaystyle\sum_{\substack{1\leq i_2 \leq [n/2]-s\\i_2 + [n/2]\leq j_2 \leq n-s}}\!\!\!\!\!\!C_{a,a,a,a,b,b}\leq \bigg(\sum_{i = 0}^\infty \Abs{b_i}\bigg)^6,
\displaystyle\sum_{\substack{1\leq i_1 \leq [n/2]-s\\i_1 + [n/2]\leq j_1 \leq n-s}}\displaystyle\sum_{\substack{1\leq i_2 \leq [n/2]-s\\i_2 + [n/2]\leq j_2 \leq n-s}}\!\!\!\!\!\!C_{a,a,a,a,a,a}\leq \bigg(\sum_{i = 0}^\infty \Abs{b_i}\bigg)^6
\end{eqnarray*}
As $\tr(\bSig)\leq \sqrt{p}\tr^{1/2}(\bSig^2)$, we have
\begin{eqnarray*}
\displaystyle\sum_{\substack{1\leq i_1 \leq [n/2]-s\\i_1 + [n/2]\leq j_1 \leq n-s}}\displaystyle\sum_{\substack{1\leq i_2 \leq [n/2]-s\\i_2 + [n/2]\leq j_2 \leq n-s}}\E(\bepsilon_{i_1}^\top\bepsilon_{j_1}\bepsilon_{i_1+r}^\top\bepsilon_{i_2}^\top\bepsilon_{j_2}\bepsilon_{i_2+r}^\top)
&=&O(n\tr^3(\bSig)+n^2\tr(\bSig)\tr(\bSig^2)+\tr^3(\bSig))\\
&=&O(np^{3/2}+n^2p^{1/2})\tr^{3/2}(\bSig^2).
\end{eqnarray*}
Since $\E(\bar{\bepsilon}_n^\top\bar{\bepsilon}_n)=O(n^{-1}p^{1/2}\tr^{1/2}(\bSig^2))$, These implies
\begin{eqnarray*}
\displaystyle\sum_{r = 1}^M\displaystyle\sum_{s = 1}^M\frac{\displaystyle\sum\nolimits_{\substack{1\leq i \leq [n/2]-s\\i + [n/2]\leq j \leq n-s}}\bepsilon_{i}^\top\bepsilon_j\bepsilon_{i+r}^\top\bar{\bepsilon}_n}{(n-s/2-\frac32 [n/2] +1/2)([n/2]-s)}=O_p(n^{-4}pM^2\tr(\Sigma^2))=o_p(\tr(\bSig^2)).
\end{eqnarray*}
Similarly, we can verify
\begin{gather*}
\displaystyle\sum_{r = 1}^M\displaystyle\sum_{s = 1}^M\frac{\displaystyle\sum\nolimits_{\substack{1\leq i \leq [n/2]-s\\i + [n/2]\leq j \leq n-s}}\bepsilon_{i}^\top\bepsilon_j\bar{\bepsilon}_n^\top\bar{\bepsilon}_n}{(n-s/2-\frac32 [n/2] +1/2)([n/2]-s)}=O_p((n^{-3/2}p^{1/2}+n^{-2}p)M\tr(\bSig^2))=o_p(\tr(\bSig^2)),\\
\displaystyle\sum_{r = 1}^M\displaystyle\sum_{s = 1}^M\frac{\displaystyle\sum\nolimits_{\substack{1\leq i \leq [n/2]-s\\i + [n/2]\leq j \leq n-s}}\bepsilon_{i}^\top\bar{\bepsilon}_n\bar{\bepsilon}_n^\top\bar{\bepsilon}_n}{(n-s/2-\frac32 [n/2] +1/2)([n/2]-s)}=O_p(n^{-2}pM^2\tr(\bSig^2))=o_p(\tr(\bSig^2)),\\
\displaystyle\sum_{r = 1}^M\displaystyle\sum_{s = 1}^M\frac{\displaystyle\sum\nolimits_{\substack{1\leq i \leq [n/2]-s\\i + [n/2]\leq j \leq n-s}}\bar{\bepsilon}_n^\top\bar{\bepsilon}_n\bar{\bepsilon}_n^\top\bar{\bepsilon}_n}{(n-s/2-\frac32 [n/2] +1/2)([n/2]-s)}=O_p(n^{-2}pM^2\tr(\bSig^2))=o_p(\tr(\bSig^2)).
\end{gather*}
\end{Prof}


\begin{thebibliography}{}

\bibitem[\protect\astroncite{Ayyala et~al.}{2017}]{ayyala2017mean}
Ayyala, D.~N., Park, J., and Roy, A. (2017).
\newblock Mean vector testing for high-dimensional dependent observations.
\newblock {\em Journal of Multivariate Analysis}, 153:136--155.

\bibitem[\protect\astroncite{Bai and Saranadasa}{1996}]{bai1996effect}
Bai, Z. and Saranadasa, H. (1996).
\newblock Effect of high dimension: by an example of a two sample problem.
\newblock {\em Statistica Sinica}, 6:311--329.

\bibitem[\protect\astroncite{Breitung and Das}{2005}]{breitung2005panel}
Breitung, J. and Das, S. (2005).
\newblock Panel unit root tests under cross-sectional dependence.
\newblock {\em Statistica Neerlandica}, 59(4):414--433.

\bibitem[\protect\astroncite{Brockwell and Davis}{2013}]{Brockwell2013TimeST}
Brockwell, P.~J. and Davis, R.~A. (2013).
\newblock Time series: Theory and methods.
\newblock Springer science \& business media.

\bibitem[\protect\astroncite{Cai et~al.}{2014}]{tony2014two}
Cai, T., Liu, W., and Xia, Y. (2014).
\newblock Two-sample test of high dimensional means under dependence.
\newblock {\em Journal of the Royal Statistical Society Series B: Statistical
  Methodology}, 76(2):349--372.

\bibitem[\protect\astroncite{Chang et~al.}{2017}]{chang2017simulation}
Chang, J., Zheng, C., Zhou, W.-X., and Zhou, W. (2017).
\newblock Simulation-based hypothesis testing of high dimensional means under
  covariance heterogeneity.
\newblock {\em Biometrics}, 73(4):1300--1310.

\bibitem[\protect\astroncite{Chen et~al.}{2024}]{chen2022asymptotic}
Chen, D.~C., Feng, L., and Liang, D.~C. (2024).
\newblock Asymptotic independence of the quadratic form and maximum of
  independent random variables with applications to high-dimensional tests.
\newblock {\em Acta Mathematica Sinica, English Series}, 40(12):3093--3126.

\bibitem[\protect\astroncite{Chen et~al.}{2011}]{chen2011regularized}
Chen, L.~S., Paul, D., Prentice, R.~L., and Wang, P. (2011).
\newblock A regularized hotelling’s ${T}^2$ test for pathway analysis in
  proteomic studies.
\newblock {\em Journal of the American Statistical Association},
  106(496):1345--1360.

\bibitem[\protect\astroncite{Chen et~al.}{2019}]{chen2019two}
Chen, S., Li, J., and Zhong, P.-S. (2019).
\newblock Two-sample and anova tests for high dimensional means.
\newblock {\em The Annals of Statistics}, 47(3):1443--1474.

\bibitem[\protect\astroncite{Chen and Qin}{2010}]{chen2010two}
Chen, S. and Qin, Y.-L. (2010).
\newblock A two-sample test for high-dimensional data with applications to
  gene-set testing.
\newblock {\em The Annals of Statistics}, 38(2):808--835.

\bibitem[\protect\astroncite{Cho et~al.}{2019}]{cho2019note}
Cho, S., Lim, J., Ayyala, D.~N., Park, J., and Roy, A. (2019).
\newblock Note on mean vector testing for high-dimensional dependent
  observations.
\newblock {\em arXiv preprint arXiv:1904.09344}.

\bibitem[\protect\astroncite{Feng et~al.}{2024}]{feng2024asymptotic}
Feng, L., Jiang, T., Li, X., and Liu, B. (2024).
\newblock Asymptotic independence of the sum and maximum of dependent random
  variables with applications to high-dimensional tests.
\newblock {\em Statistica Sinica}, 34:1745--1763.

\bibitem[\protect\astroncite{Feng et~al.}{2022}]{feng2022testing}
Feng, L., Liu, B., and Ma, Y. (2022).
\newblock Testing for high-dimensional white noise.
\newblock {\em arXiv preprint arXiv:2211.02964}.

\bibitem[\protect\astroncite{Feng et~al.}{2016}]{feng2016multivariate}
Feng, L., Zou, C., and Wang, Z. (2016).
\newblock Multivariate-sign-based high-dimensional tests for the two-sample
  location problem.
\newblock {\em Journal of the American Statistical Association},
  111(514):721--735.

\bibitem[\protect\astroncite{Gregory et~al.}{2015}]{gregory2015two}
Gregory, K.~B., Carroll, R.~J., Baladandayuthapani, V., and Lahiri, S.~N.
  (2015).
\newblock A two-sample test for equality of means in high dimension.
\newblock {\em Journal of the American Statistical Association},
  110(510):837--849.

\bibitem[\protect\astroncite{He et~al.}{2021}]{he2021asymptotically}
He, Y., Xu, G., Wu, C., and Pan, W. (2021).
\newblock Asymptotically independent u-statistics in high-dimensional testing.
\newblock {\em Annals of statistics}, 49(1):154.

\bibitem[\protect\astroncite{Huang et~al.}{2022}]{huang2022overview}
Huang, Y., Li, C., Li, R., and Yang, S. (2022).
\newblock An overview of tests on high-dimensional means.
\newblock {\em Journal of Multivariate Analysis}, 188:104813.

\bibitem[\protect\astroncite{Li et~al.}{2020}]{li2020adaptable}
Li, H., Aue, A., Paul, D., Peng, J., and Wang, P. (2020).
\newblock An adaptable generalization of hotelling’st $^2$ test in high
  dimension.
\newblock {\em The Annals of Statistics}, 48(3):1815--1847.

\bibitem[\protect\astroncite{Li et~al.}{2016}]{li2016simpler}
Li, Y., Wang, Z., and Zou, C. (2016).
\newblock A simpler spatial-sign-based two-sample test for high-dimensional
  data.
\newblock {\em Journal of Multivariate Analysis}, 149:192--198.

\bibitem[\protect\astroncite{Pollard}{2012}]{Pollard1984ConvergenceOS}
Pollard, D. (2012).
\newblock Convergence of stochastic processes.
\newblock Springer Science \& Business Media.

\bibitem[\protect\astroncite{Srivastava}{2009}]{srivastava2009test}
Srivastava, M.~S. (2009).
\newblock A test for the mean vector with fewer observations than the dimension
  under non-normality.
\newblock {\em Journal of Multivariate Analysis}, 100(3):518--532.

\bibitem[\protect\astroncite{Srivastava and Du}{2008}]{srivastava2008test}
Srivastava, M.~S. and Du, M. (2008).
\newblock A test for the mean vector with fewer observations than the
  dimension.
\newblock {\em Journal of Multivariate Analysis}, 99(3):386--402.

\bibitem[\protect\astroncite{Srivastava et~al.}{2016}]{srivastava2016raptt}
Srivastava, R., Li, P., and Ruppert, D. (2016).
\newblock Raptt: An exact two-sample test in high dimensions using random
  projections.
\newblock {\em Journal of Computational and Graphical Statistics},
  25(3):954--970.

\bibitem[\protect\astroncite{Thulin}{2014}]{thulin2014high}
Thulin, M. (2014).
\newblock A high-dimensional two-sample test for the mean using random
  subspaces.
\newblock {\em Computational Statistics $\&$ Data Analysis}, 74:26--38.

\bibitem[\protect\astroncite{Wang et~al.}{2015}]{wang2015high}
Wang, L., Peng, B., and Li, R. (2015).
\newblock A high-dimensional nonparametric multivariate test for mean vector.
\newblock {\em Journal of the American Statistical Association},
  110(512):1658--1669.

\bibitem[\protect\astroncite{Xu et~al.}{2016}]{xu2016adaptive}
Xu, G., Lin, L., Wei, P., and Pan, W. (2016).
\newblock An adaptive two-sample test for high-dimensional means.
\newblock {\em Biometrika}, 103(3):609--624.

\bibitem[\protect\astroncite{Xue and Yao}{2020}]{xue2020distribution}
Xue, K. and Yao, F. (2020).
\newblock Distribution and correlation-free two-sample test of high-dimensional
  means.
\newblock {\em The Annals of Statistics}, 48(3):1304--1328.

\bibitem[\protect\astroncite{Zhang et~al.}{2018}]{zhang2018clt}
Zhang, B., Pan, G., and Gao, J. (2018).
\newblock {CLT} for largest eigenvalues and unit root testing for
  high-dimensional nonstationary time series.
\newblock {\em The Annals of Statistics}, 46(5):2186--2215.

\bibitem[\protect\astroncite{Zhang et~al.}{2025}]{zhang2023MeanTF}
Zhang, S., Chen, S.~X., and Qiu, Y. (2025).
\newblock Mean tests for high-dimensional time series.
\newblock {\em Statistica Sinica}, 35(1):171--201.

\bibitem[\protect\astroncite{Zhong et~al.}{2013}]{zhong2013tests}
Zhong, P.-S., Chen, S.~X., and Xu, M. (2013).
\newblock Tests alternative to higher criticism for high-dimensional means
  under sparsity and column-wise dependence.
\newblock {\em The Annals of Statistics}, 41(6):2820--2851.

\end{thebibliography}
\end{document}